\documentclass[onecolumn, journal, letterpaper,12pt]{IEEEtran}

\usepackage{graphicx, epstopdf}
\usepackage{xr-hyper}
\usepackage{amssymb, amsmath}
\usepackage{commath}
\usepackage{setspace}
\usepackage{acronym}
\usepackage{mathrsfs}
\usepackage{ifthen}
\usepackage{textcomp}
\usepackage{float}
\usepackage{subfig}
\usepackage[table]{xcolor}
\usepackage{color,soul}
\usepackage{cite}
\usepackage{setspace}
\usepackage{mathtools}
\usepackage{bm}
\usepackage{url}
\usepackage{algorithm}
\usepackage{algpseudocode}
\usepackage[switch]{lineno}
\usepackage[english]{babel}
\usepackage[utf8]{inputenc}
\usepackage{amsmath,amsthm}
\usepackage{amssymb}
\usepackage{dsfont}
\usepackage{grffile}
\usepackage{subfig}
\usepackage{tikz}
\usepackage{authblk}
\usepackage[colorinlistoftodos]{todonotes}
\theoremstyle{plain}

\theoremstyle{definition}

\theoremstyle{remark}

\usepackage{algorithm}
\usepackage{algpseudocode}
\usepackage{enumitem}
\usepackage{float}
\usepackage{mathtools}
\usepackage{url}
\usetikzlibrary{calc}
\usepackage{hyperref}
\usepackage{url}
\usepackage{lineno}

\IEEEoverridecommandlockouts
\begin{document}

\title{Impact of Geographic Diversity on Citation of Collaborative Research}  

\author[1]{\small Cian Naik}
\author[2]{\small Cassidy R. Sugimoto}
\author[3]{\small Vincent Larivi{\`e}re}
\author[4,6]{\small Chenlei Leng}
\author[5,6,*]{\small Weisi Guo}

\affil[1]{\footnotesize University of Oxford, United Kingdom}
\affil[2]{\footnotesize Georgia Institue of Technology, USA}
\affil[3]{\footnotesize University of Montreal, Canada}
\affil[4]{\footnotesize University of Warwick, United Kingdom}
\affil[5]{\footnotesize Cranfield University, United Kingdom}
\affil[6]{\footnotesize Alan Turing Institute, United Kingdom}
\affil[*]{\footnotesize Corresponding author: Weisi Guo, weisi.guo@cranfield.ac.uk}

\markboth{current preprint version (v1) APRIL 2022}
{Submitted paper}

\maketitle

\begin{abstract}
Diversity in human capital is widely seen as critical to creating holistic and high quality research, especially in areas that engage with diverse cultures, environments, and challenges. Quantifying diverse academic collaborations and its effect on research quality is lacking, especially at international scale and across different domains. Here, we present the first effort to measure the impact of geographic diversity in coauthorships on the citation of their papers across different academic domains. Our results unequivocally show that geographic coauthor diversity improves paper citation, but very long distance collaborations has variable impact. We also discover "well-trodden" collaboration circles that yield much less impact than similar travel distances. These relationships are observed to exist across different subject areas, but with varying strengths. These findings can help academics identify new opportunities from a diversity perspective, as well as inform funders on areas that require additional mobility support.
\end{abstract}

\section{Introduction} \label{introduction}
International collaboration is a key part of scientific research, with the exchange of ideas from diverse sources leading to numerous breakthroughs. A recent paper by \cite{sugimoto2017scientists} showed that researchers with affiliations to more than one country during their career, so-called "mobile" researchers, had a significant boost in citations over their non-mobile colleagues. Indeed, several well established international initiatives (Marie Curie Staff Exchange, German DAAD, Royal Society International Exchange) fund researcher mobility between countries and across disciplines. An important facilitator in long-distance collaboration is the ease of air transportation between locations. 

\subsection{Relevant Research}

Collaboration in science is not new. Despite being often seen as a contemporary practice, research collaboration has always existed—although many collaborators were invisible from the authors’ lists \cite{shapin1989invisible}. Already in the early 19th Century, a scientist like Einstein - who is wrongly seen as a ``lone genius'' — was collaborating with colleagues on many aspects of his research \cite{janssen2015history,pyenson1985young}. The first discipline to exhibit collaboration in the form of co-authorship was chemistry. Already in 1900, 34\% of papers in the field had more than one author, compared with 10\% in physics and less than 1\% in mathematics \cite{gingras2010transformation}. 

After the second world war, the large influx of research funding and the era of ``big science'' has led to an important rise in collaboration activities and, as consequence, of multi-authored papers \cite{wuchty2007increasing}. Since the beginning of the 1950s, most papers have more than one author in the natural and medical sciences \cite{cronin2003cast,franceschet2010effect,galison2003collective,persson2004inflationary,wuchty2007increasing}, while single authorship remained the norm in social sciences and humanities until the early 2000s \cite{lariviere2015team}. In the latter group of disciplines, social sciences and arts and humanities have distinct practices: while the majority of papers in social sciences are the results of collaboration, single authorship remains the norm in arts and humanities \cite{larivi2006canadian}. At the other end of the spectrum, fields such as high energy physics have author lists that have gone beyond 5000 names, a phenomenon named hyper-authorship \cite{cronin2005hand}. Such decline in single authorship has had long been predicted \cite{price1986little}, and shown empirically in the work of Harriet Zuckerman \cite{zuckerman1967nobel}. Indeed, focusing on Nobel Laureates between 1900 and 1959, she shows that after 1920, most of the laureates’ papers are the result of collaboration. The rise in collaborative activities can also be linked with an increase in international collaboration \cite{sonnenwald2007scientific, wagner2005network}, which is also observed in all fields but the arts and humanities \cite{larivi2006canadian}. Such growth is observed both in terms of the share of papers that are in international collaboration, as well as the number of countries involved \cite{lariviere2015team}. 

\subsubsection{Multi-Faceted Nature of Collaboration}
Several factors can be associated with this rise in researchers’ collaborative activities. The first factor is the ease with which technology allowed researchers to communicate and conduct research \cite{katz1997research}. Since the advent of the digital age, technologies such as the Internet, email and online communication platforms such as Skype, Zoom, or Teams have allowed researchers to exchange data, meet, and write papers at a distance with much more ease than what was previously possible. Despite those technologies, previous research shows that there remains an effect of distance, where researchers are more likely to collaborate with colleagues that are physically closer \cite{abramo2009research,catalini2018microgeography,gieryn2002three,hoekman2010research}. Another factor is its epistemic effect—i.e. its effect on scientific impact \cite{wray2002epistemic}. Science is increasingly complex, and larger teams are therefore necessary to tackle contemporary scientific problems. This has been shown empirically, as collaborative research is associated with higher citation rates \cite{franceschet2010effect,narin1991scientific,wuchty2007increasing}. This is specifically true for international collaboration \cite{glanzel2001national}. This can also be associated with infrastructure: big science infrastructures have become so expensive that they have to be shared, often internationally. This is particularly true for smaller countries \cite{luukkonen1992understanding}. This positive relationship has been observed already in the early 20th century \cite{lariviere2015team}. A third factor is policies from funders and universities. Indeed, some countries have made policies that emphasized collaboration, especially international \cite{abramo2009research}) or interdisciplinary \cite{NAP11153}. Such policies are based on the fact that countries’ resources are limited, and that collaboration is considered to lead to more important scientific results. A fourth factor is specialization: in a context where researchers are increasingly specialized, collaboration allows for researchers with complementary expertise to work together on a research problem \cite{franceschet2010effect}. 

\subsubsection{Importance of Distance \& Diversity}
Despite the importance of digital technology in making long-distance collaboration possible, in person collaborations are still conducted. In this context, the possibility of traveling between two cities can be hypothesized to have an effect on the likelihood of collaboration, and reduce the effect of physical distance. Previous analyses \cite{ploszaj2020impact} have been performed, using data on flight capacity and frequency, as well as collaboration. Using a sample of four universities in the United States, they have shown that more flights between cities and the proximity of airports to universities are linked with higher numbers of collaborations. Unsurprisingly, collaboration was higher in cases where direct flights can be obtained between the cites. \cite{catalini2020travel} also show that not only does travel cost constitute a friction to collaboration, a reduction to this friction leads to a increase in higher-quality projects. However, air travel is not necessarily associated with academic success. Research in \cite{wynes2019academic} has shown, using a sample of researchers from the University of British Columbia (Canada) that, once controlling for age and discipline, air travel emissions were not associated with higher impact measures, although traveling was associated with higher salaries. Recent work at university level by \cite{guo2017global} showed that the connectivity of universities via the air transport network is an important indicator of ranking growth for the universities, even after accounting for economic development.

\begin{figure}[htb]
\centering
\includegraphics[width=0.95\textwidth]{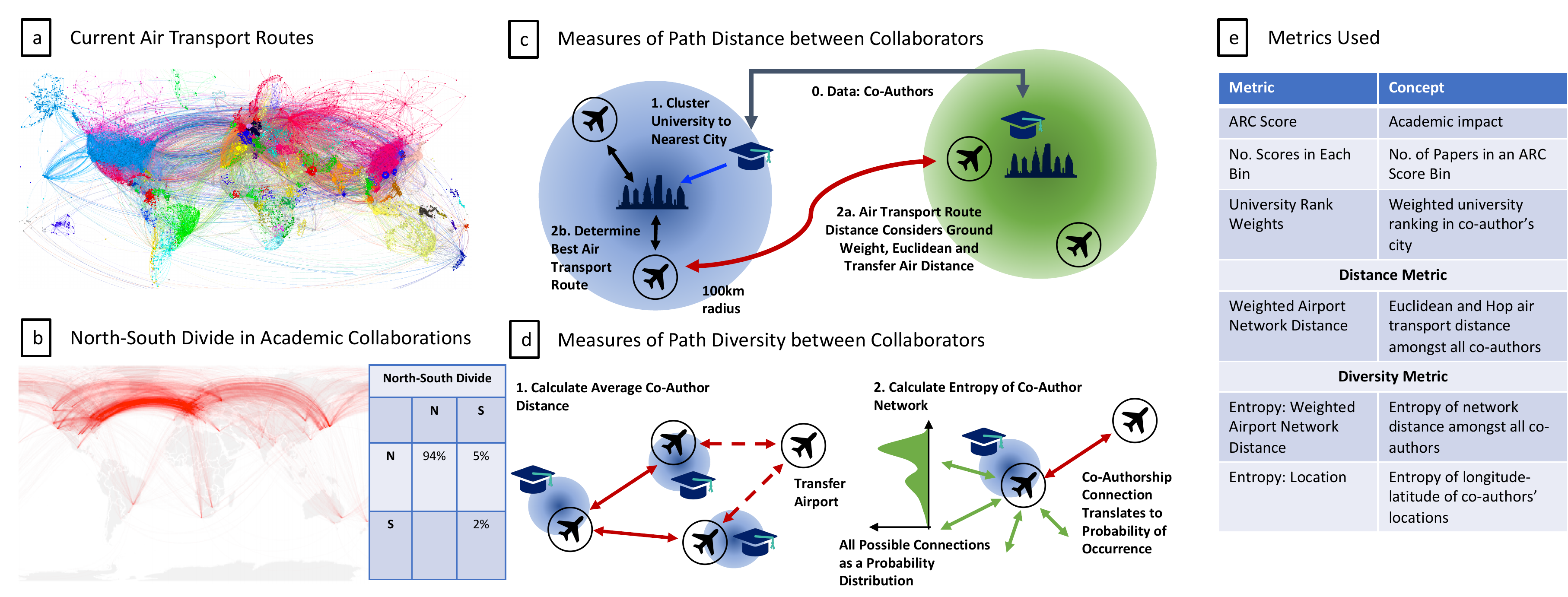}
\caption{Diversity analysis of coauthorship networks. In $(a)$ we plot the global flight connections. $(b)$ gives the corresponding plot for a selection of academic collaborations. $(c)$ introduces the factors that compose our distance metric. $(d)$ introduces the corresponding factors for the diversity metrics. $(e)$ lists the metrics we use.}                          
\label{fig:1}  
\end{figure}

\subsection{Contribution}
Building on these ideas, we use the air transport network to quantify the geographical diversity in paper coauthorships. The air transport network is a network of connections between cities (\textit{nodes}) where the \textit{edges} are flights. We use it to define measures of diversity between the researchers based in these cities, with full details provided on how we do this in the methods section. We focus on establishing a link between the geographical diversity of coauthors on a given paper and the number of citations that paper receives. As shown in Figure~\ref{fig:1}, a novelty is to develop distance and entropy measures for diversity on the coauthorship network and evaluate the variation of the Average Relative Citation (ARC) score against these.

The rest of the paper is structured as follows. In Section \ref{results} we present the key results. In Section \ref{stats_results}, we present the robustness of our results to potential confounding variables, such as the effect of university rankings. In Section \ref{comparison_results} we examine the results by subject area and location, in order to examine subject and geographic specific differences. We provide details of the data and methods we use for this analysis in Section \ref{methods}. We discuss implications on individual academics, universities, funders, and government policy in Section \ref{sec:impact}. In the Appendix, we include some additional results.

\begin{figure}[htb]
\centering
\includegraphics[width=0.95\textwidth]{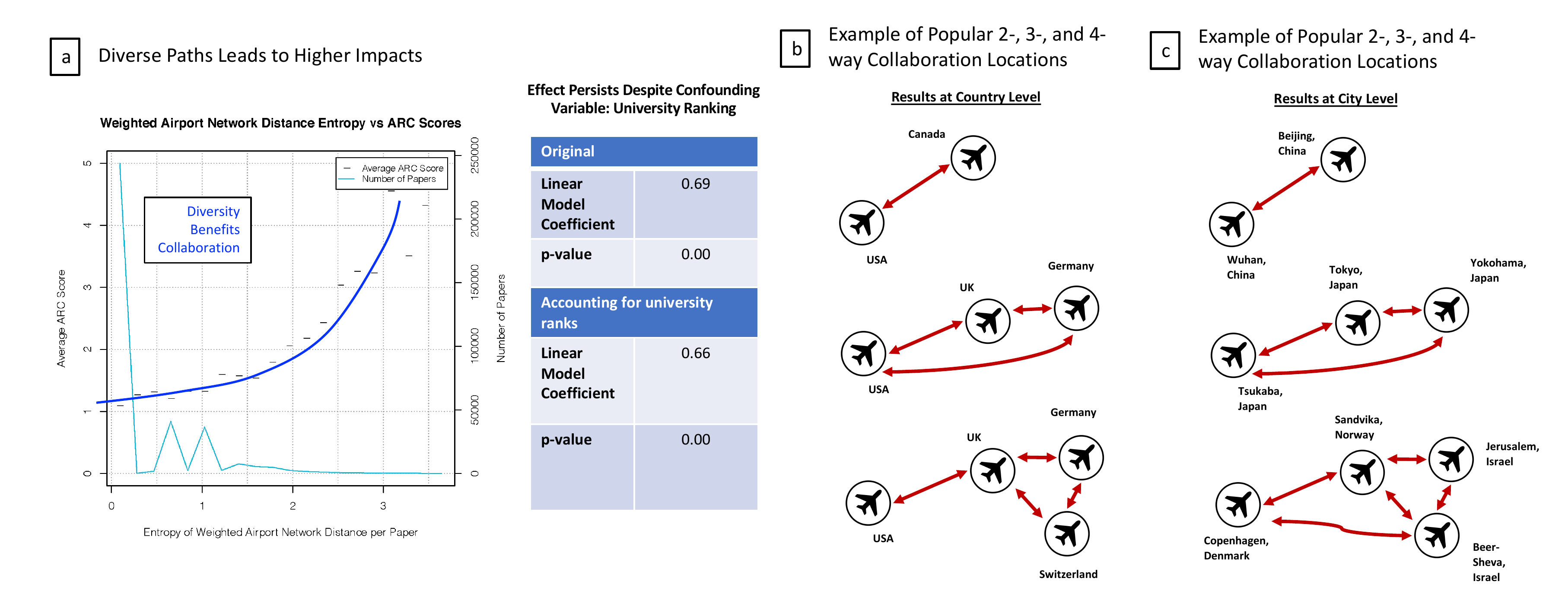}
\caption{Headline results showing that diverse collaborations lead to greater ARC. $(a)$ shows the relationship between weighted airport network distance entropy and average ARC score. $(b)$ and $(c)$ give examples of popular collaboration routes at the country and city level respectively.}                          
\label{fig:2}  
\end{figure}

\begin{figure}[htb]
\centering
\includegraphics[width=0.95\textwidth]{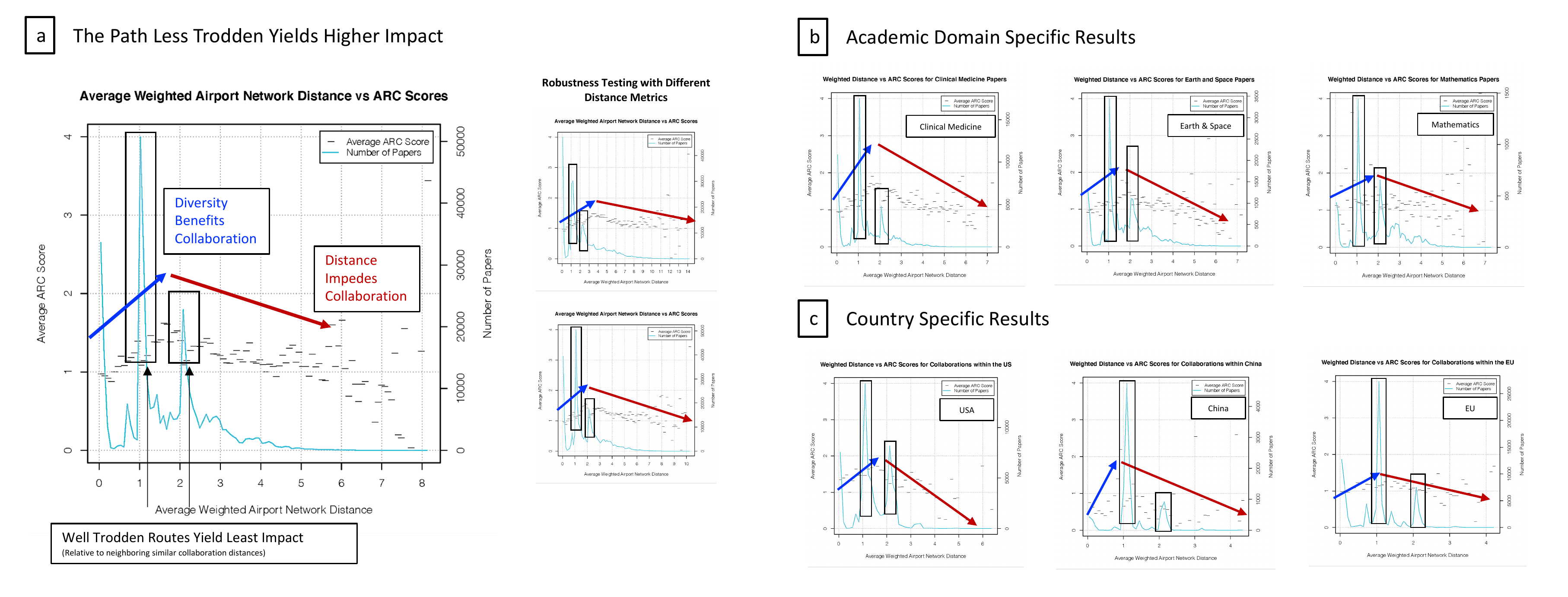}
\caption{Relationship between average weighted airport network distance and average ARC score, showing that well trodden paths and extreme long distance collaborations can reduce ARC. In $(a)$ we look at the overall relationship, before breaking it down by $(b)$ academic domain and $(c)$ country.}                          
\label{fig:3}  
\end{figure}

\section{Results} \label{results}

\subsection{Main Discoveries}

\textbf{Diverse Collaborations Lead to Higher Citations.} Our primary main discovery is that for a relatively simple notion of diversity measured by the entropy of the probability of forming a collaboration, the ARC score is highly correlated with the entropy, as seen in Figure~\ref{fig:2}(a). We are aware of certain confounding variables, chiefly the potential effect that university rankings has on citations \cite{clauset2015systematic}. We show that this correlation persists even when accounting for this. We also reveal some popular "well-trodden" 2-, 3-, and 4-way collaboration paths in Figures~\ref{fig:2}(b)-(c).

\textbf{Well-Trodden Paths and Extreme Distances Lead to Relatively Lower Citations.} Our secondary main discovery is that the aforementioned "well-trodden" paths yield relatively lower citation than similar distances and that extremely long distance collaborations have variable or reduced citation. Using the air transport network distance metric, we show in Figure~\ref{fig:3}(a) how diversity initially benefits collaboration until distance takes its toll and impedes frequent exchange of ideas. Local spikes in the number of collaborations exist in the general data set, specific academic domains, and specific countries. These spikes correspond to well-trodden collaboration paths - see Figures~\ref{fig:2}(b)-(c) (highlighted by a black box in Figure~\ref{fig:3}) also correspond to local "dips" in ARC scores. That is to say, well-trodden collaboration paths do not yield as much citation as similar distances between other collaboration locations. We observe this pattern across all domains and countries, but note exaggerated effects in certain cases, e.g., long distance collaboration is more detrimental in clinical medicine (possibly due to the practical and timely nature of its practice).

\textbf{A North-South Divide Exists in Collaborative Research.} Finally, our third main discovery is that a divide exists in the composition of collaborative research, with most collaborations occurring between researchers located in the global north. When looking at pairs of collaborations (where a collaboration between more than two authors contains multiple pairs), we see from Figure~\ref{fig:1}(b) that 94\% of collaboration pairs are between researchers in the global north.

\begin{figure}[htb]
\centering
\subfloat[Average weighted airport network distance \label{fig:distance_mean}]{\includegraphics[width=0.4\textwidth]{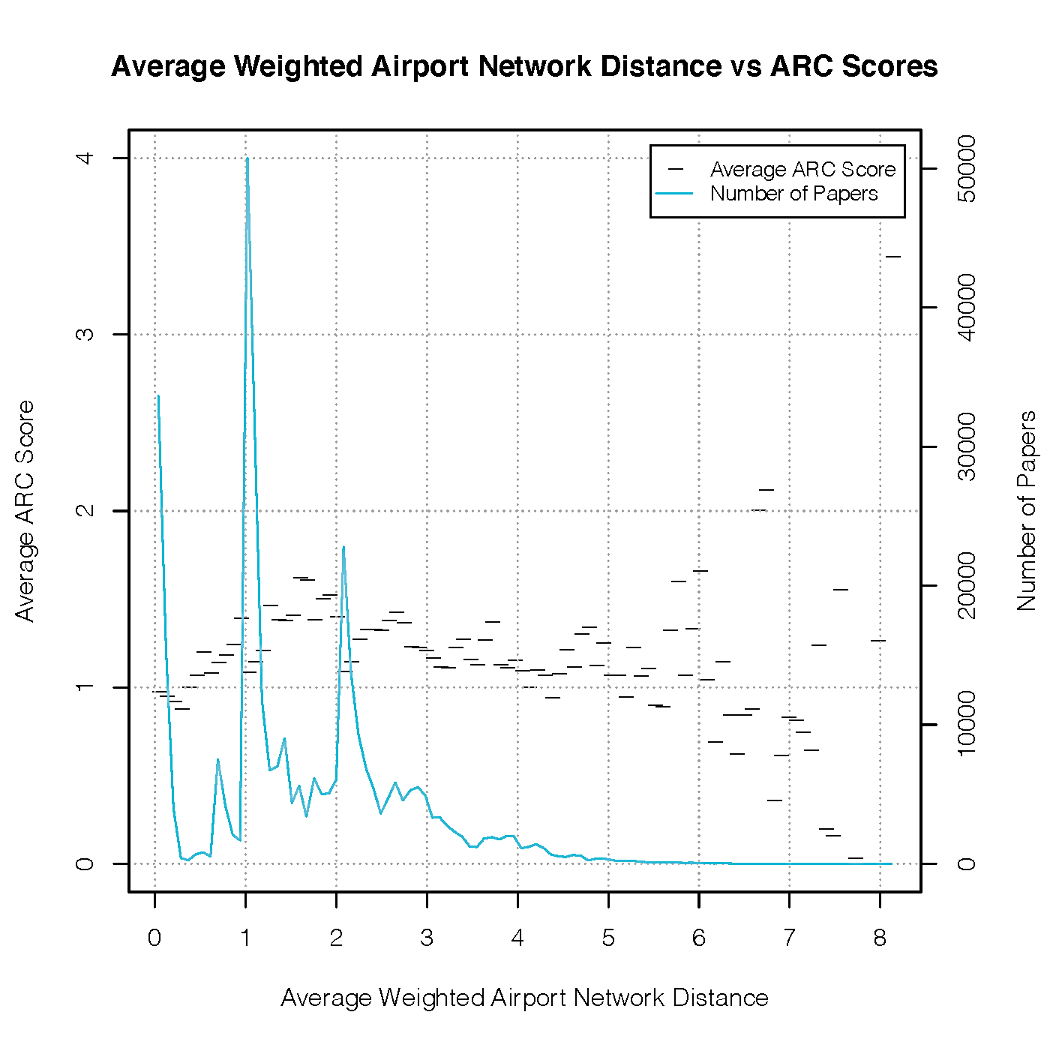}}
\subfloat[Weighted airport network distance entropy \label{fig:distance_entropy}]{\includegraphics[width=0.4\textwidth]{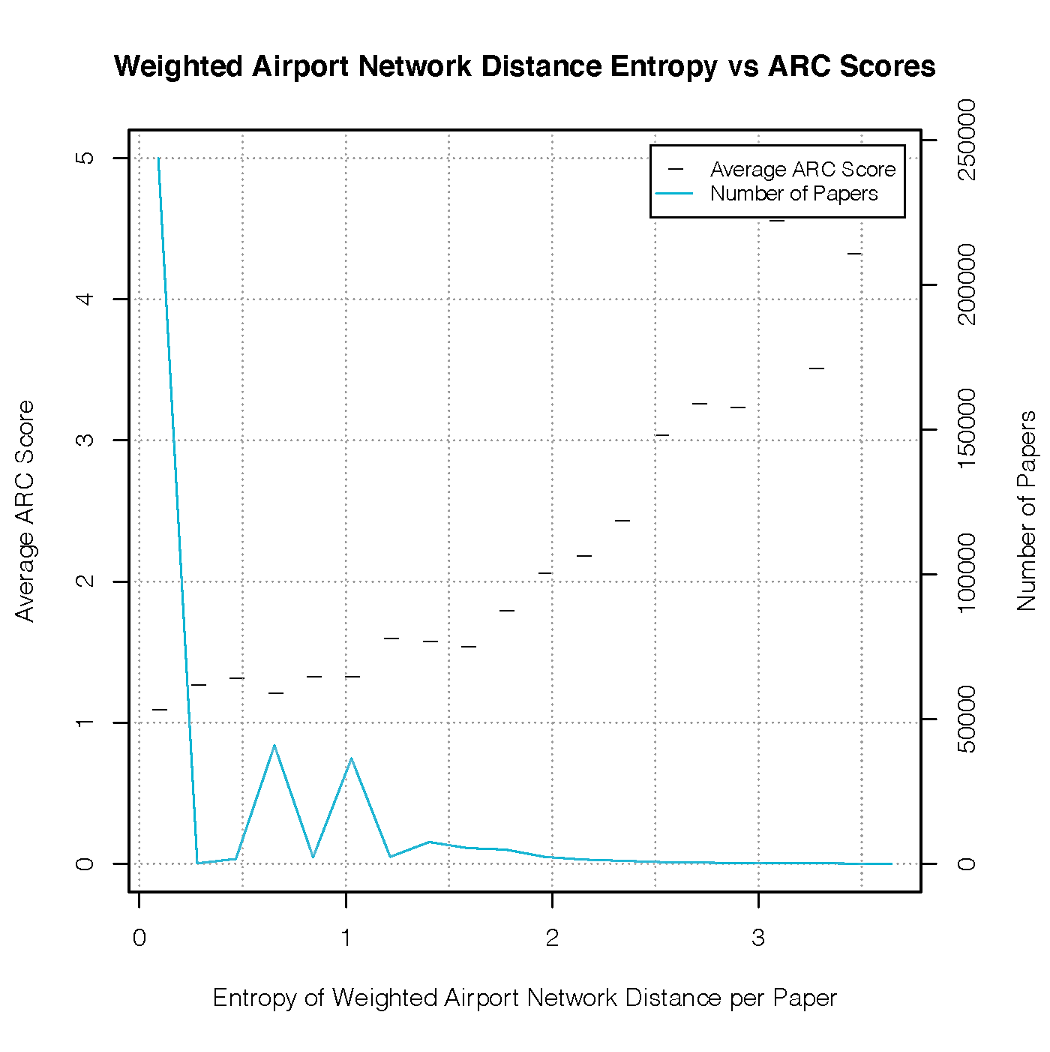}}\\
\subfloat[Weighted entropy of coauthor location\label{fig:weighted_entropy}]{\includegraphics[width=0.4\textwidth]{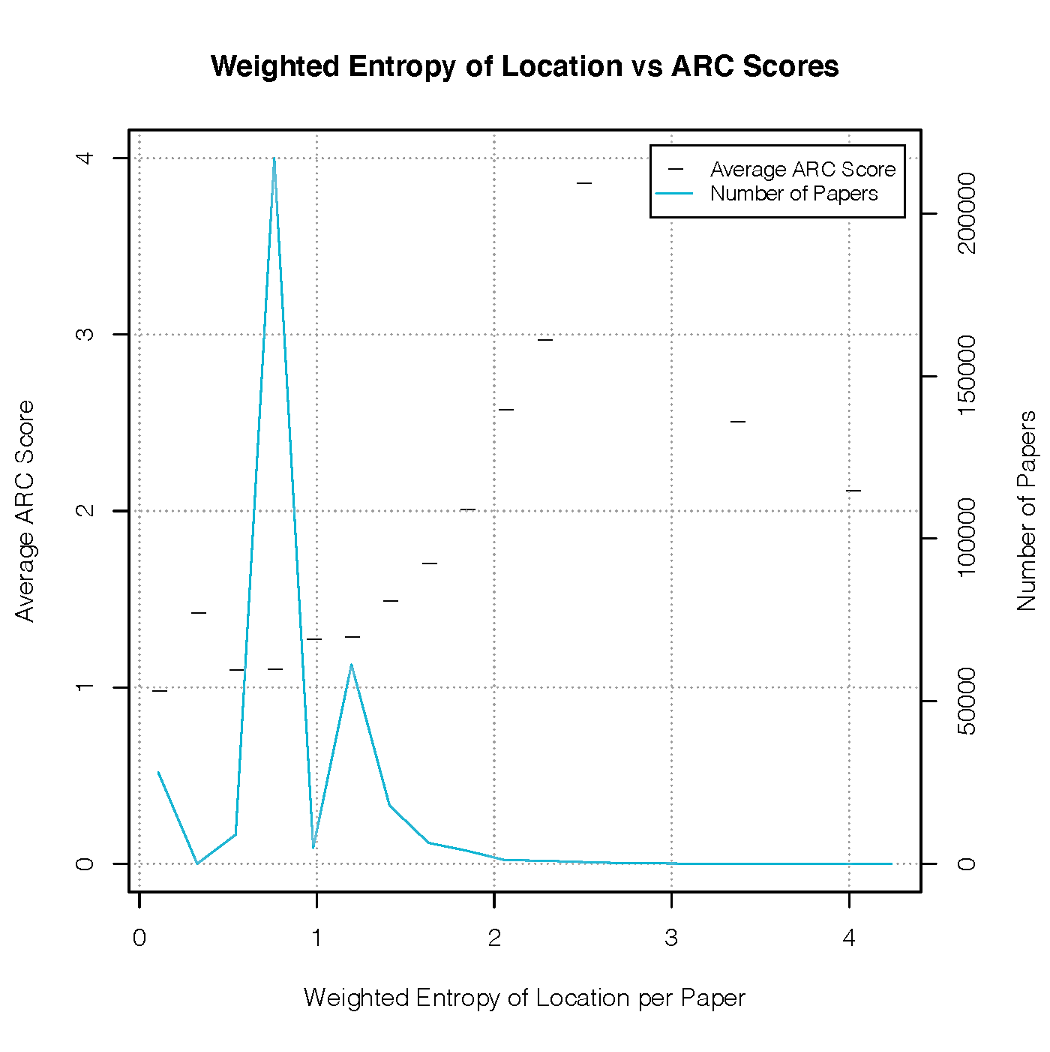}}
\subfloat[Average university rank weight\label{fig:rank_mean}]{\includegraphics[width=0.4\textwidth]{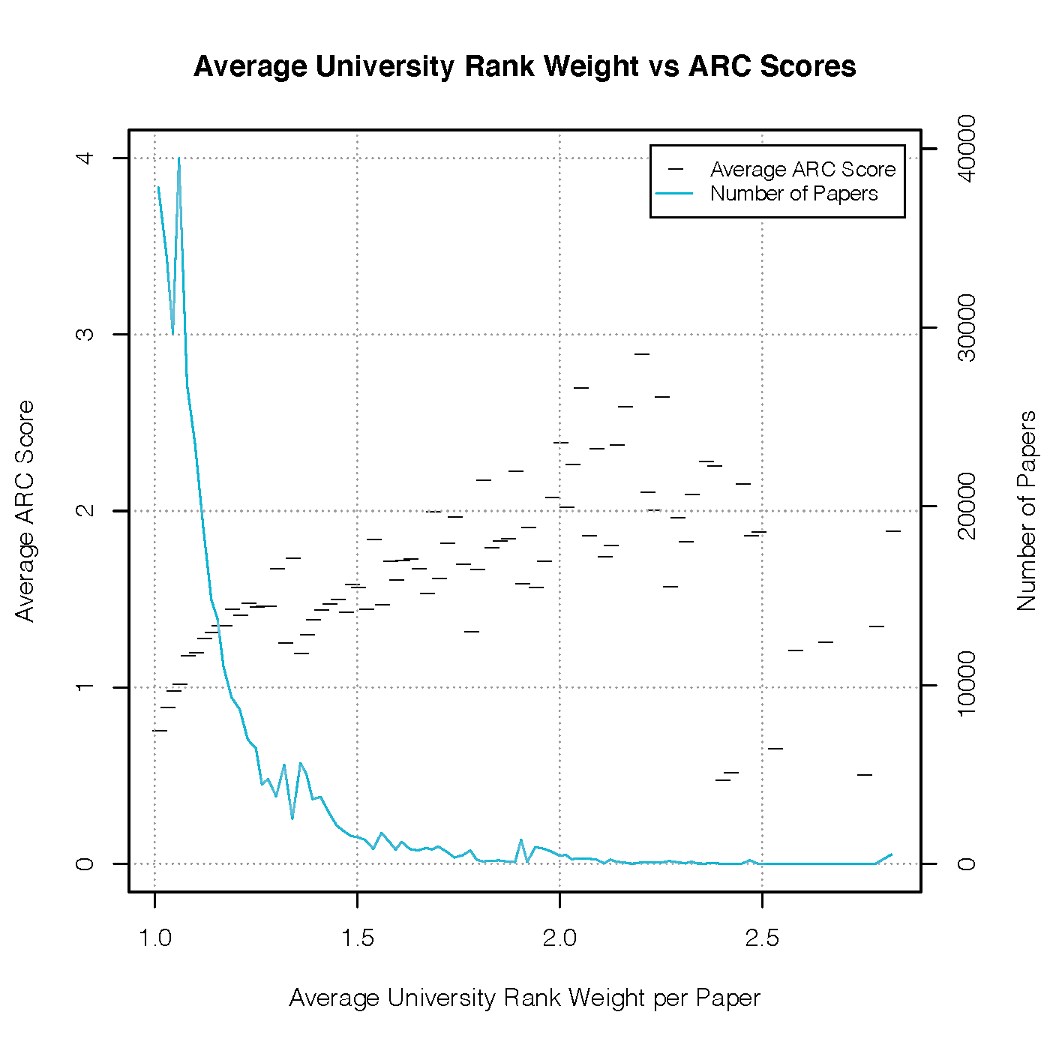}}
\caption{Binned comparison of $(a)$ average weighted airport network distance, $(b)$ weighted airport network distance entropy, $(c)$ weighted entropy of coauthor location, and $(d)$ average university rank weight against ARC Score.}   
\label{initial_results}  
\end{figure}

\subsection{Detailed Analysis of Effect of Distance, Diversity, and University Rank on ARC Scores}
In Figure \ref{fig:1}(e), we briefly introduce four important measures whose relationship with ARC scores we are interested in investigating. We give a more detailed explanation of these here, with the full derivation of the measures presented in Section \ref{methods}. We also identify some key patterns we see in the relationships with ARC score, which can be seen in Figure \ref{initial_results}.
\begin{enumerate}
\item \textbf{[Collaboration Distance] Average Weighted Airport Network Distance}. This is a measure of the average distance between collaborators on a given paper. The distance is the weighted network distance on the flight network. Based on the work of \cite{gastner2006spatial}, an edge on the network is assigned a weight
\begin{align} \label{eq:euc_weights}
\text{effective length of edge } (i,j) = \lambda d_{ij}+(1-\lambda) 
\end{align}
where $d_{ij}$ is the Euclidean distance between nodes $i$ and $j$, and $\lambda$ is a parameter that controls the importance of physical distance against graph distance. From Figure \ref{fig:distance_mean}, we see a positive correlation between citations and this measure of distance. However, past a certain point, we see that the number of citations decreases. We can conjecture that the large average distance could mean that these coauthors are in remote areas, geographically and in terms of transport links. 
\item \textbf{[Collaboration Diversity] Weighted Airport Network Distance Entropy}. This measure also looks at the weighted network distance between coauthors. It uses a more direct measure of diversity - the entropy of these distances. In Figure \ref{fig:distance_entropy} we see that as this measure of diversity increases, the number of citations also increases consistently, showing a clear trend between diversity and citations.
\item \textbf{[Alternative Collaboration Diversity] Weighted Entropy of Coauthor Location}. In this alternative measure of diversity, we consider the entropy of the geographic locations of the coauthors. In this case a weighted entropy measure is used (not to be confused with the weighted distances introduced previously). The ``weight'' in this case incorporates the centrality of node on the flight network, as well as university rankings. Again we see in Figure \ref{fig:weighted_entropy} that as this measure of diversity increases, the number of citations also increases consistently, showing a clear trend between diversity and citations.
\item \textbf{[Important Confounding Factor] Average University Rank Weight}. This measure weights cities by the average world ranking of the universities located within a certain radius. This is important to consider, as the reputation of a university can have a significant effect on the number citations received by papers produced by its researchers \cite{clauset2015systematic}. In Figure \ref{fig:rank_mean} we see a strong correlation between the university rank weights and number of citations. This effect seems to flatten out somewhat as the average weight increases. This could be indicating that the effect of university rankings is less important for the top universities. However, it could also come from our specific choice of the construction of the weights. The exact nature of this relationship is outside the scope of this work.
\end{enumerate}

In each of the plots comprising Figure \ref{initial_results} the data is binned. In each case, we also plot the number of papers that are in each bin. In addition to the main results already presented, we see that the variability of the Average ARC score increases for large values of each of these measures. We can see that these cases correspond to a very small number of papers, and so this is not unexpected. 

\subsection{Robustness of Results to Parameter Choices and Confounding Variables}\label{parameter_robustness_counding}
There are two key situations in which we check the robustness of the results obtained. The first of these concerns the key configuration parameter $\lambda$, which controls the balance between Euclidean distance and flight hop distance in (\ref{eq:euc_weights}).  In our case, we choose a value of $\lambda=\frac{1}{10000}$, as this gives some interpretability which we lose for larger choices, as detailed in Section \ref{methods}. However, the results we observe can also be seen for different choices of $\lambda$. One exception to this is that for much larger choices such as $\lambda=\frac{1}{5}$, the weighted distances are completely dominated by the Euclidean distances. In this case we lose the interpretation of ``Well-trodden paths''. Further discussion is presented in the Appendix.

Secondly, as noted above, it is well known that there is a strong link between university rankings and paper citations \cite{clauset2015systematic}. The relationship of interest in our case is therefore the effect that our distance and diversity measures have on ARC score, specifically not occuring via university rankings (since this is a relationship that is already well understood). In order to disentangle these effects, we explicitly account for the confounding effects of unversity rankings. We see that the patterns already observed still persist having done so. In Section \ref{stats_results} we present the full analysis controlling for this effect. In particular, the results displayed in Tables \ref{table:comp_conf} and \ref{table:ent_comp} give evidence to support our claims.

\section{Statistical Analysis of Results}\label{stats_results}
So far we have presented results which have been largely qualitative in nature. We have observed two distinct trends in the average ARC score with increasing average distance and entropy of distance between coauthors. However, we now wish to quantify these results. Motivated by the patterns of the points in Figure \ref{fig:distance_mean}, we first define a model to check for the existence, location and significance of the ``peak'' we observe in the relationship between average weighted network distance and ARC score.
\subsection{Average Weighted Airport Network Distance}
\begin{figure}[htb]
\centering
\includegraphics[width=0.6\textwidth]{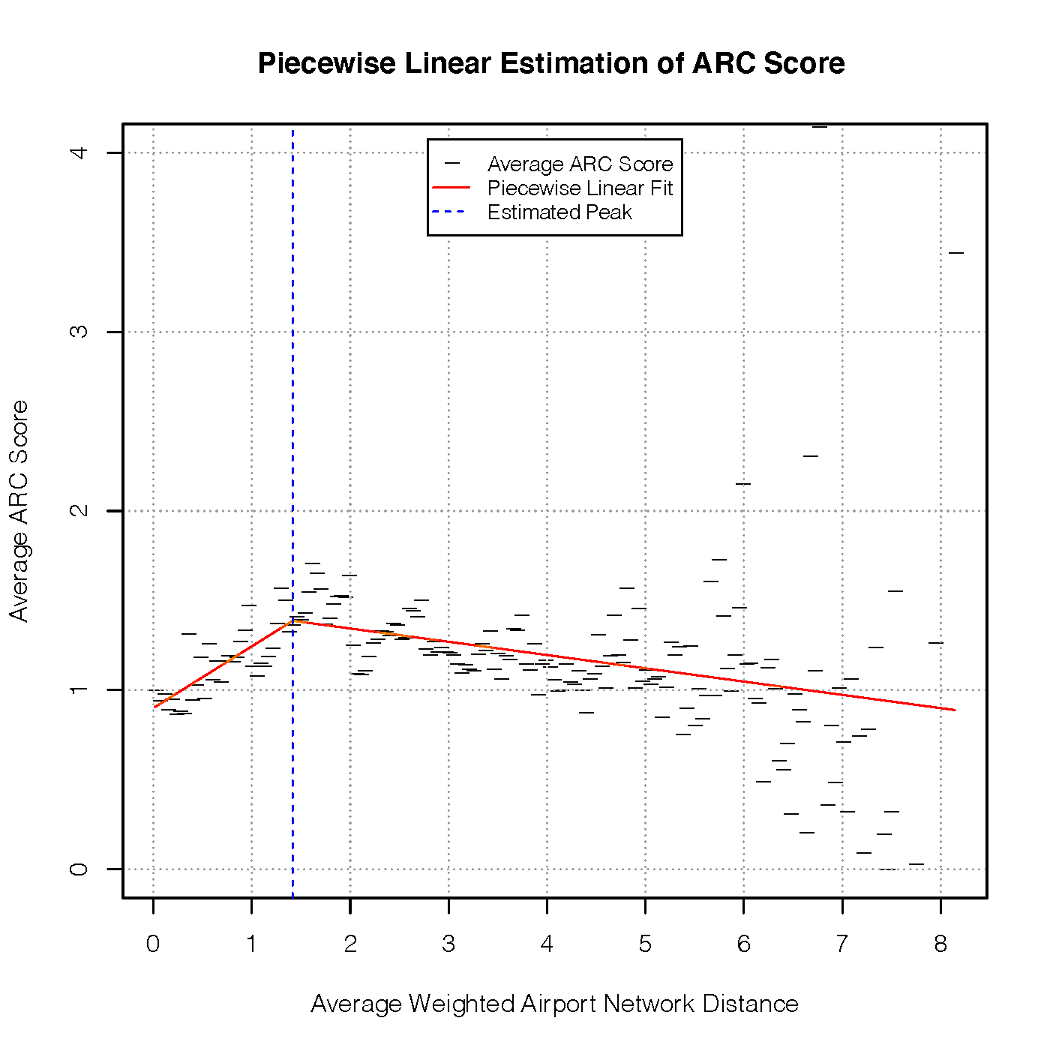}
\caption{Piecewise linear estimation of the relationship between average weighted distance and ARC score}                                    
\label{fig:pl_original}  
\end{figure}

In order to check for the existence and location of a peak, we fit a piecewise linear model, limited to two pieces. The model can be summarised as:
\begin{align}\label{eq:piecewise}
    f(x) = \begin{cases}
    a_1 + b_1x &x\leq x^*\\
    a_2 + b_2x &x>x^*
\end{cases}
\end{align}
where $a_1, b_1, a_2, b_2$ are such that $f(x)$ is continuous at $x^*$. The model is fitted for a range of values $x^*$, and is optimized to find the value of $x^*$ for which the residual sum of squares is lowest. The optimal value $\hat{x}^*$ gives the estimated location of the peak. We can test whether a statistically significant peak exists by checking that the corresponding gradients $\hat{b}_1, \hat{b}_2$ are significantly $\geq 0$ and $\leq 0$ respectively\footnote{In this case we define significance at the $5\%$ level by checking that the p-values are $\leq 0.05$}. In Figure \ref{fig:pl_original} we see an example of what this fit looks like. Our analysis confirms what we intuitively saw in Figure \ref{fig:distance_mean}, with a statistically significant increase and decrease in average ARC before and after the peak\footnote{Throughout our analysis, we fit the piecewise linear model on the raw (rather than binned) data, but for ease of understanding we show the fit on the binned plot. However, in practice we find that the results are very similar if we perform a weighted fit to the binned data using the number of data points in each bin.}. We emphasise that our goal here is not to accurately model the relationship that we observe, but merely to confirm the existence of this peaked shape that we see in the data. For this purpose, a simple piecewise linear model works well. More complicated models may capture the relationship better, but that is outside the scope of this work.

\begin{figure}[htb]
\centering
\includegraphics[width=0.4\textwidth]{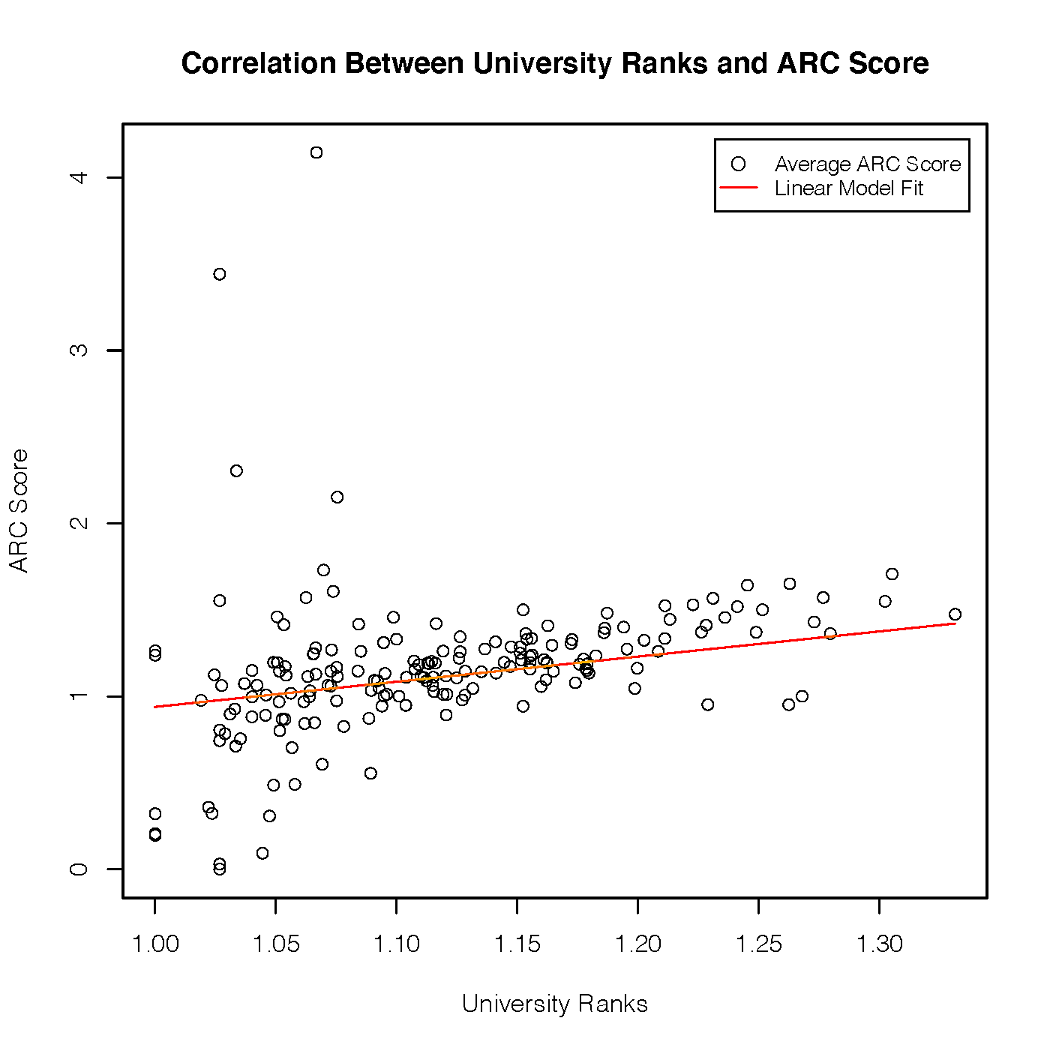}
\caption{Correlation between university rank score and ARC score}                                    
\label{fig:rank_arc_cor}  
\end{figure}

This does not yet tell the full story. As before we can test for the pattern detailed above after removing the effect of university rankings, as mentioned in Section \ref{parameter_robustness_counding}. The effect that they have on citations received by papers is already well studied \cite{clauset2015systematic}. We can see this clearly if we plot the (binned) university rank weights (as defined in (\ref{eq:uni_rank_scores})) against the average ARC scores. We do this in Figure \ref{fig:rank_arc_cor} and see an almost linear relationship. 

Disentangling how much of the relationship between average weighted distance and ARC score occurs via university ranks is a potentially difficult task, and we do not focus on that in our work. Instead, we take a conservative approach, removing as much of the effect of university ranks as possible by directly fitting ARC score against average university rank weights, and removing that effect before fitting the piecewise linear model of ARC score against average weighted distance. Specifically, letting $y_{ARC}$ be the ARC score for each paper, $d_{AV}$ be the average weighted airport network distance between the coauthors and $w_{AV}$ the average university rank weights of the coauthor locations, we first estimate $\hat{y}_{ARC}$ from $y_{ARC} \sim w_{AV}$. Then we fit our piecewise model $y_{ARC} - \hat{y}_{ARC} \sim f(d_{AV})$, where $f(x)$ is defined as in (\ref{eq:piecewise}). 

We compare the unadjusted fit (as seen in Figure \ref{fig:pl_original}) with the corresponding fit having adjusted for the effect of the university ranks in this way, with the results given in Table \ref{table:comp_conf}. We see that the observed increase stays almost constant, as does the peak location. However, the decrease that we observe seems to be at least partly tied in the the University ranks.

\begin{table}[ht]
\centering
\begin{tabular}{|r||rrrrr|}
  \hline
 Method & $\hat{x}^*$ & $\hat{b}_1$ & p-value & $\hat{b}_2$  & p-value \\ 
  \hline
Before adjusting & 1.60 & 0.25 & 0.00 & -0.08 & 0.00 \\ 
After adjusting & 1.65 & 0.24 & 0.00 & -0.04 & 0.00 \\ 
\hline
\end{tabular}
\caption{Fitting a piecewise linear model for ARC score using average weighted airport network distance, before and after adjusting for the effect of university rankings}                                                         
\label{table:comp_conf}
\end{table}
Further analysis is presented in the Appendix, where we use stratification to support the results presented here.

\subsection{Weighted Airport Network Distance Entropy}
We now investigate the relationship between weighted airport network distance entropy and ARC score. In Figure \ref{fig:distance_entropy} we see that the average ARC score increases as the entropy increases. In order to test whether this increase is significant, the first step is to fit a linear model of ARC score against weighted distance entropy, having accounted for university rankings. Specifically, letting $y_{ARC}$ be the ARC score for each paper, $d_{ENT}$ be the average weighted airport network distance between the coauthors and $w_{AV}$ the average university rank weights of the coauthor locations, we first estimate $\hat{y}_{ARC}$ from $y_{ARC} \sim w_{AV}$. Then we fit the simple model $y_{ARC} - \hat{y}_{ARC} \sim d_{ENT}$. Again, we emphasise that our goal here is not to accurately model the relationship that we observe, and that other models may provide a better fit than the linear model that we use. However, our goal is simply to confirm the existence of a statistically significant trend.

\begin{figure}[htb]
\centering
\includegraphics[width=0.6\textwidth]{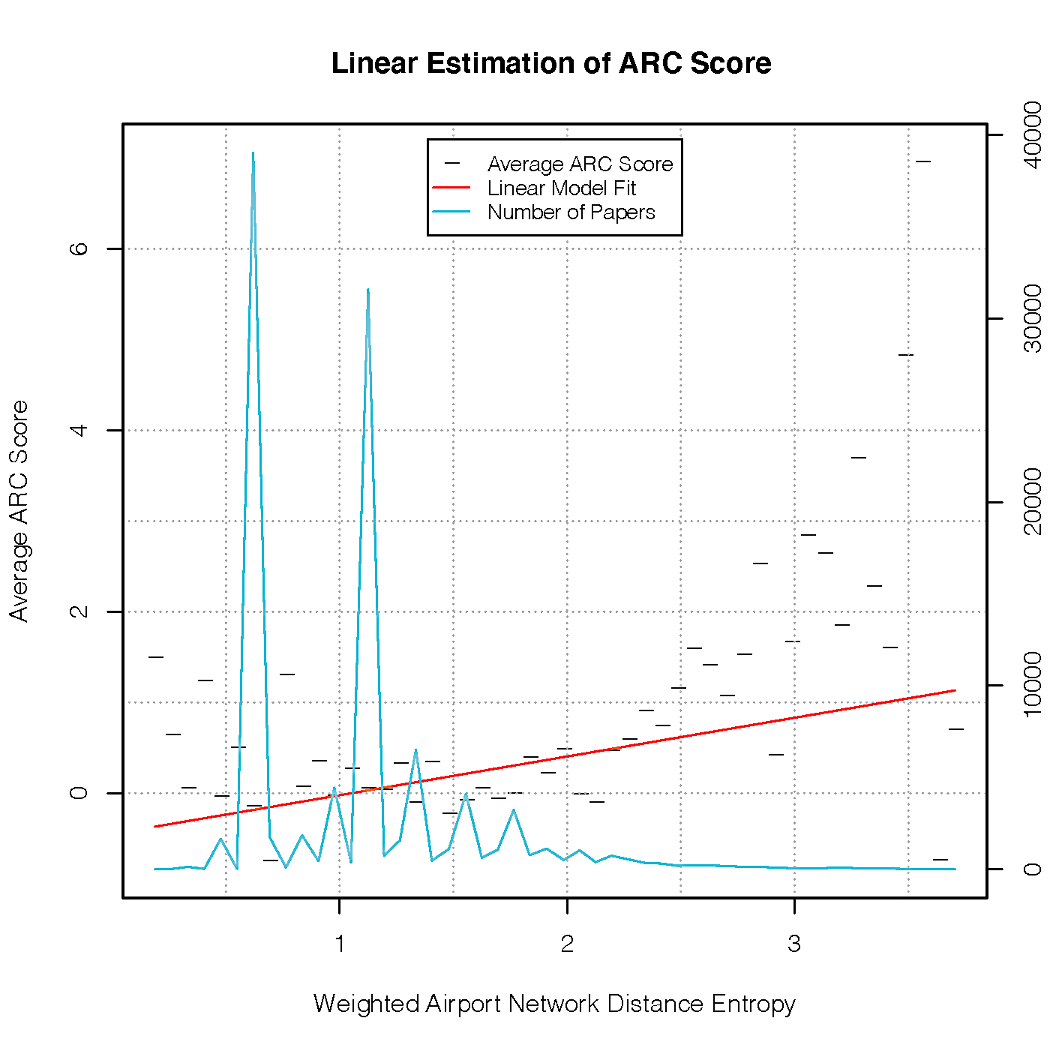}
\caption{Linear estimation of ARC score using weighted distance entropy}
\label{fig:ent_overall}  
\end{figure}

In Table \ref{table:ent_comp} we see the estimated parameters from fitting the above model, and from fitting the model without adjusting for university rankings. In each case we see a significant increase in ARC score as distance entropy increases. In Figure \ref{fig:ent_overall} we see the fit of the model, having accounted for university rankings. A linear model does not capture the behaviour of the data as well as the piece-wise linear model fit for the average weighted distance metric. In fact, it looks as though the average ARC scores initially decrease as the entropy increases. The reason for this is that we fit the model with the full data, but plot the binned data. As we can see from the numbers of papers in each bin, most of the bins have very few values, and the model fit is dominated by the two large spikes. Thus, in Figure \ref{fig:ent_overall}, the higher ARC scores for very small values of the distance entropy are somewhat misleading, as are the corresponding results for very large values of the distance entropy.

\begin{table}[ht]
\centering
\begin{tabular}{|r||rr|}
  \hline
 Method & $\hat{b}$ & p-value \\ 
  \hline
Before adjusting & 0.69 & 0.00  \\ 
After adjusting & 0.66 & 0.00  \\ 
\hline
\end{tabular}
\caption{Fitting a linear model for ARC score using weighted airport network distance entropy, before and after adjusting for the effect of university rankings}                                                         
\label{table:ent_comp}
\end{table}

\section{Comparisons}\label{comparison_results}
Having defined methods to analyse our results quantitatively, and to control for the effect of university rankings, we now break the overall results down by academic field and coauthor location, in order to gain a better insight into the trends that are occurring.

\subsection{Results by Academic Field} 
\subsubsection{Average Weighted Airport Network Distance} Fistly, we compare different fields based on the location of the peak in the relationship between average weighted network distance and ARC score. We also compare the gradients before and after, to see how prominent the peak is. In Table \ref{table:podr} we see the results. There are several interesting features we notice here. Firstly, we see that for all the fields but one, there is a significant positive relationship until a point. Secondly, we notice that we can broadly split the different fields into three different categories, based on the patterns exhibited:
\begin{enumerate}
    \item Fields such as  Social Sciences, Clinical Medicine and Biomedical research, which exhibit the peaked form described earlier, with significant increases and decreases. 
    \item Fields such as Physics, Engineering and technology and Psychology, which exhibit a significant initial positive relationship, but subsequently plateau, with no significant positive or negative relationship.
    \item Mathematics, which does not seem to exhibit any significant relationship.
\end{enumerate}
Lastly, if we examine the point at which there is no longer a positive relationship (either the peak or the start of the plateau) then we see differences between the field. In Table \ref{table:podr} we have sorted the fields by the estimate of $\hat{x}^*$, and we see that for fields such as Biology and Psychology increasing the average weighted network distance has a positive effect on ARC scores for much longer than for fields such as Social Sciences and Engineering and Technology.
\begin{table}[ht]
\centering
\begin{tabular}{|r||rrrrr|}
  \hline
 Field & $\hat{x}^*$ & $\hat{b}_1$ & p-value & $\hat{b}_2$  & p-value \\ 
  \hline
Social Sciences & 1.37 & 0.38 & 0.00 & -0.10 & 0.01 \\ 
Engineering and Technology & 1.43 & 0.26 & 0.00 & 0.01 & 0.64 \\ 
Professional Fields & 1.46 & 0.46 & 0.00 & -0.15 & 0.00 \\ 
Clinical Medicine & 1.65 & 0.34 & 0.00 & -0.10 & 0.00 \\ 
Physics & 1.65 & 0.21 & 0.00 & -0.01 & 0.79 \\ 
Health & 1.67 & 0.27 & 0.00 & -0.07 & 0.33 \\ 
Biomedical Research & 1.69 & 0.25 & 0.00 & -0.06 & 0.00 \\ 
Chemistry & 1.76 & 0.11 & 0.00 & 0.04 & 0.13 \\ 
Earth and Space & 1.86 & 0.25 & 0.00 & -0.09 & 0.00 \\ 
Psychology & 1.90 & 0.22 & 0.00 & -0.01 & 0.75 \\ 
Biology & 2.72 & 0.07 & 0.00 & -0.01 & 0.54 \\ 
Mathematics & 3.96 & 0.01 & 0.65 & 0.17 & 0.19 \\ 
\hline
\end{tabular}
\caption{Comparison of relationships between average weighted network distance and ARC score for different fields}                                                         
\label{table:podr}
\end{table}
\subsubsection{Weighted Airport Network Distance Entropy}
We can perform the same comparison for the weighted distance entropy measure. In this case, we rank the subjects based on their estimated coefficients. We see from Table \ref{table:podr_ent} that whilst the positive relationship between entropy and ARC score exists for every subject considered, the strength of that relationship varies greatly. Mathematics and Chemistry exhibit a much weaker relationship than the other subjects, whilst Social Sciences and Clinical Medicine exhibit the strongest relationship. An important factor to consider here is the number of coauthors that papers in each field generally have. This measure of diversity only makes sense for papers with more than two coauthors, but we know that medical papers can sometimes have very large numbers of authors, whilst mathematics papers often have only a handful. It may be valuable to examine further how this factor impacts the differing relationships we see here.
\begin{table}[ht]
\centering
\begin{tabular}{|r||rr|}
  \hline
 Field & $\hat{b}$ & p-value  \\ 
  \hline
Mathematics & 0.15 & 0.00 \\ 
  Chemistry & 0.18 & 0.00 \\ 
  Psychology & 0.26 & 0.00 \\ 
  Professional.Fields & 0.28 & 0.00 \\ 
  Biology & 0.29 & 0.00 \\ 
  Physics & 0.29 & 0.00 \\ 
  Engineering.and.Technology & 0.30 & 0.00 \\ 
  Health & 0.30 & 0.00 \\ 
  Earth.and.Space & 0.35 & 0.00 \\ 
  Biomedical.Research & 0.38 & 0.00 \\ 
  Social.Sciences & 0.43 & 0.00 \\ 
  Clinical.Medicine & 0.56 & 0.00 \\ 
   \hline
\end{tabular}

\caption{Comparison of relationships between weighted network distance entropy and ARC score for different fields}                                                         
\label{table:podr_ent}
\end{table}

\subsection{Results by City}\label{sec:city_results}
\begin{figure}[htb]
\centering
\subfloat[Beijing                                 
\label{fig:beijing_piecewise}]{\includegraphics[width=0.3\textwidth]{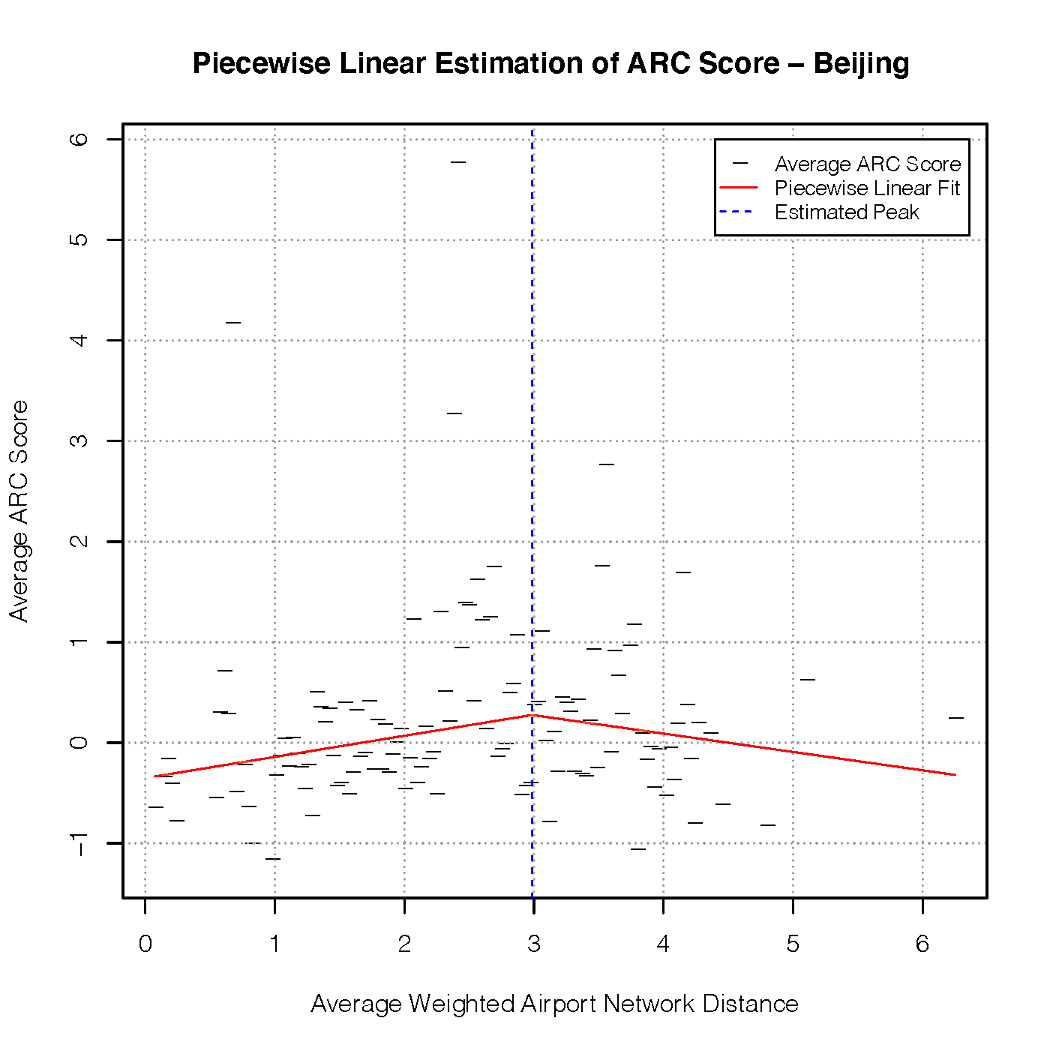}}  
\subfloat[Boston                              
\label{fig:boston_piecewise}]{\includegraphics[width=0.3\textwidth]{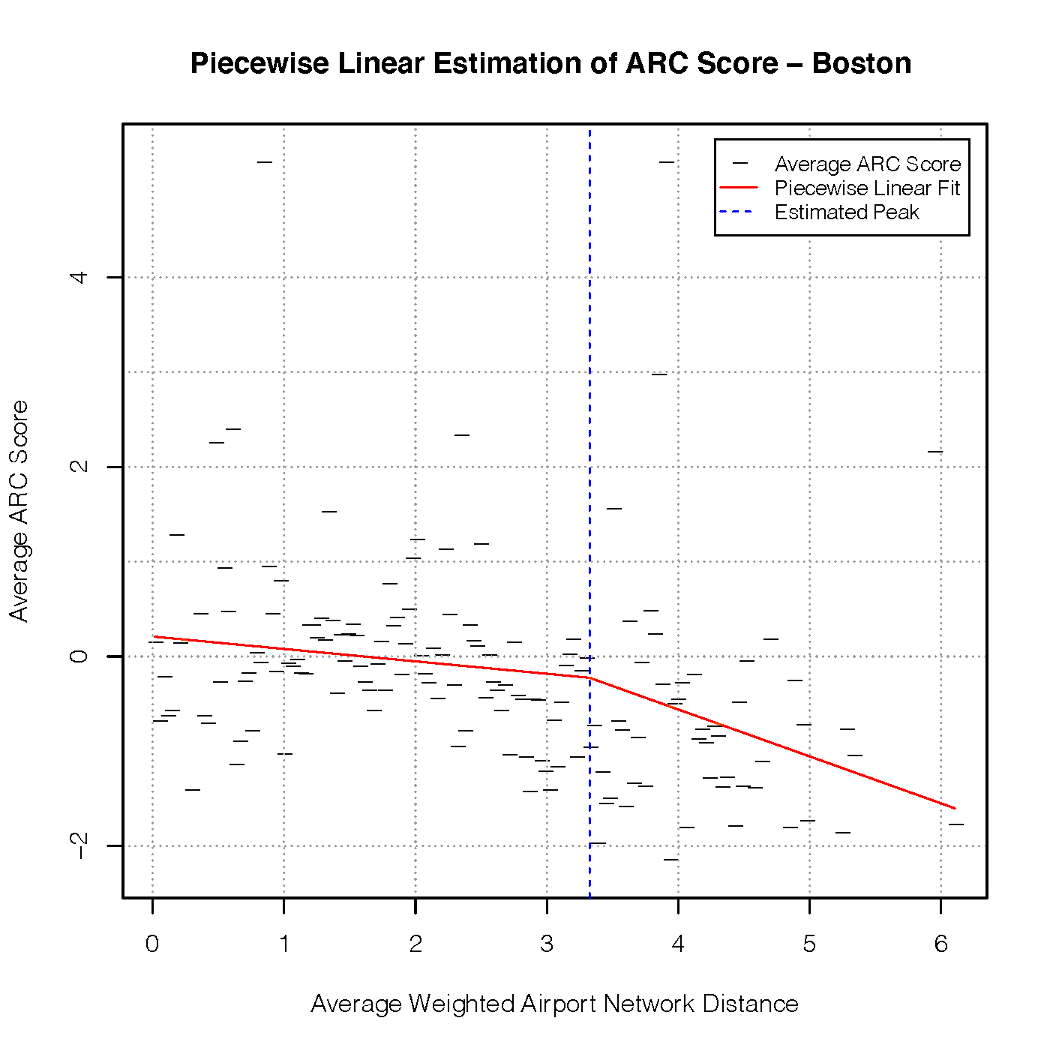}}
\subfloat[London                              
\label{fig:london_piecewise}]{\includegraphics[width=0.3\textwidth]{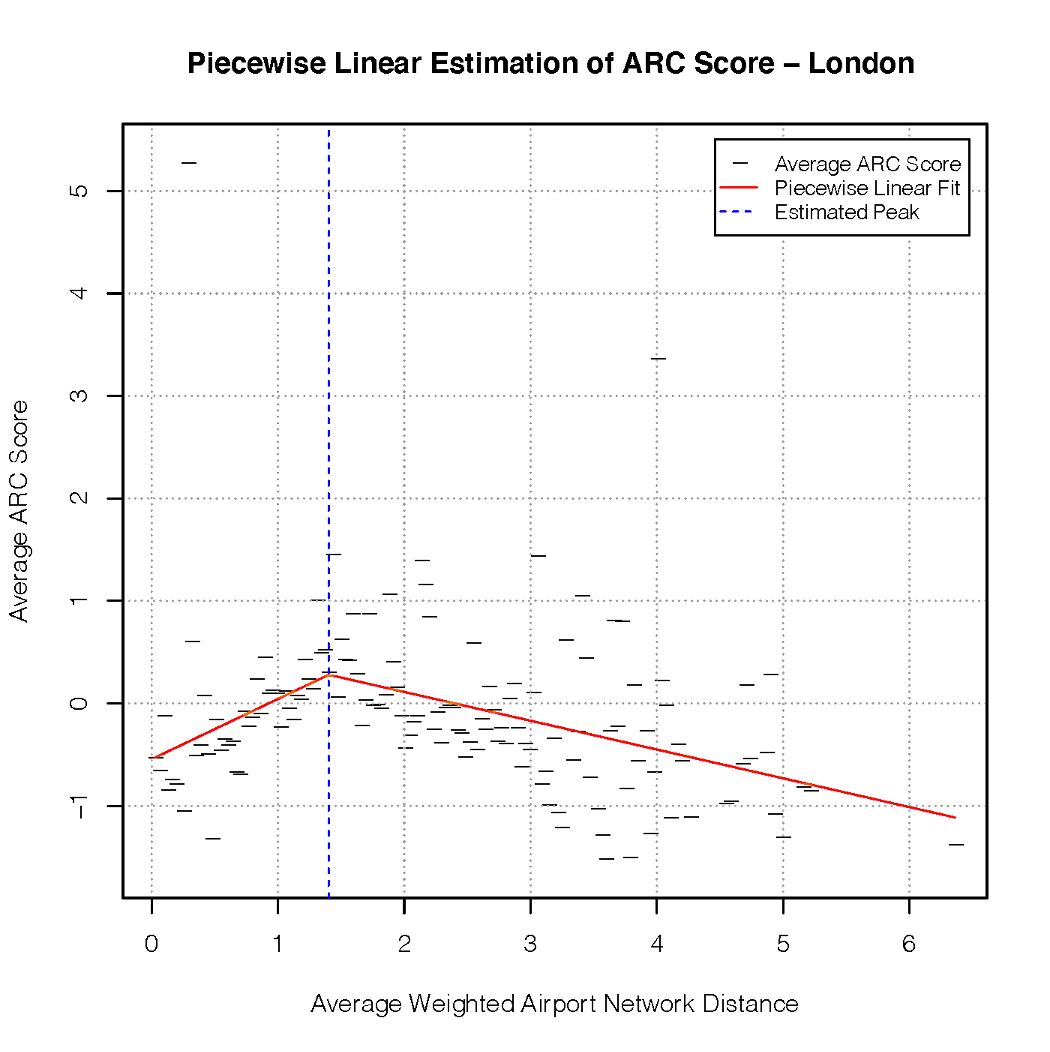}}
\caption{Piecewise linear estimation of ARC score using average weighted airport network distance, for $(a)$ Beijing, $(b)$ Boston and $(c)$ London}             
\label{fig:city_graphs}  
\end{figure}

Secondly, we compare the collaborations involving certain cities, in order to investigate differences in the collaboration patterns of their researchers. In Figure \ref{fig:beijing_piecewise} we see the plot of average weighted network distance against ARC score for Beijing, with Figures \ref{fig:boston_piecewise} and \ref{fig:london_piecewise} showing the results for Boston and London respectively. The three patterns we can see are noticeably different. For Beijing and London, there are clear peaks, but the peak for London occurs at less than half that of Beijing. Meanwhile, for Boston, it appears that there is no peak at all. A closer examination reveals that while there does still appear to be a peaked relationship, some collaborations only a small distance away from Boston but with very high ARC scores are distorting this result. 

This is certainly interesting, in terms of understanding how these cities collaborate with others. However, a slight complication arises when comparing cities in this way. Although we can see three distinct patterns here, it is not yet clear how much of these differences arise from fundamentally different behaviours of the researchers in these cities, and how much is simply due to the geographies of the cities. For example, we might expect that the most productive collaborations for researchers from Beijing are those with large American centres of research, which would generally be a weighted network distance of $2-3$ away. Similarly, for researchers from London, the weighted network distances to major European and American centres of research will be roughly between $1.2$ and $1.9$. Finally, the highly productive collaborations that researchers from Boston have are often from nearby Cambridge (home to Harvard and M.I.T.), or other East-coast cities with large research institutions.

In order to try and reduce these geographical effects, we can compare cities where we imagine that the geographical effects would be similar. We see some of these comparisons in Table \ref{table:case_studies}. From this we can see that even between cities with similar geographical effects, there can be a significant difference in the observed patterns, especially with regards to the magnitude of the initial positive effect that increasing diversity has.

\begin{table}[ht]
\centering
\begin{tabular}{|r||rrrrr|}
  \hline
 City & $\hat{x}^*$ & $\hat{b}_1$ & p-value & $\hat{b}_2$  & p-value \\ 
  \hline
Boston & 3.32 & -0.13 & 0.02 & -0.50 & 0.11 \\ 
Cambridge (USA) & 0.84 & 0.43 & 0.20 & -0.23 & 0.03 \\ 
New York & 0.90 & 0.74 & 0.00 & -0.41 & 0.00 \\ 
Berkeley & 1.30 & 0.68 & 0.00 & -0.20 & 0.10 \\ 
\hline
London & 1.40 & 0.58 & 0.00 & -0.28 & 0.00 \\ 
Oxford & 1.62 & 0.31 & 0.02 & -0.20 & 0.15 \\ 
Edinburgh & 1.98 & 0.62 & 0.00 & -0.52 & 0.00 \\ 
Dublin & 1.43 & 0.82 & 0.02 & -0.19 & 0.20 \\
\hline
Beijing & 2.96 & 0.21 & 0.00 & -0.18 & 0.57 \\ 
Hong Kong & 2.42 & 0.27 & 0.02 & -0.24 & 0.33 \\ 
\hline
\end{tabular}
\caption{Comparison of relationships between average weighted network distance and ARC score for different cities}                                                         
\label{table:case_studies}
\end{table}

\subsection{Further Work}
In this work, we focus on testing whether there is a significant increase in the average ARC score as the entropy measures increase, rather than measuring this effect. Similarly, for the average weighted airport network distance, we look for the existence and location of a peak using a piecewise linear model, without considering how well this model fits the data. While in each case these models are suitable for our purposes, further work would be needed to more accurately model the relationships we observe.

Thus far, we have also been using fairly simple models to control for the effect of university rankings. In order to better understand the results, we may want to fit more complicated models by accounting for possible nonlinear effects of the variables involved. We may also want to investigate other factors that may affect ARC scores apart from university ranks, such as economic development. 

Finally, our work has been looking at one specific year of data. An interesting extension would be to investigate if the relationships we have found differ for different years, and if so try to measure how the changing pattern of airline travel corresponds to the change in collaboration patterns.

\section{Methods}\label{methods}
Here we detail the data and methods that we use in our analysis. In particular, in Section \ref{data} we describe the data and in Section \ref{analysis} we detail how the measures of diversity that we use are constructed. 

\subsection{Data} \label{data}

\subsubsection{Coauthorship Network}
This network consists of collaborations between different coauthors, where for each collaboration we have the location of each coauthor, an identifier for the paper, and a citation score for the paper. The citation score relates to the number of citations the paper received, normalized based on the subject area. This is called the Average Relative Citation (ARC) score. The data consists of 352,057 papers published in 2005, with coauthors from 21,131 different locations. The locations of the coauthors are given as cities rather than universities, This means that we need to construct a mapping from universities to cities in order to incorporate university rankings into our analysis, as we shall describe. 

\subsubsection{Air Transport Network}
We take a snapshot of the air transport network in 2005 as a representative network showing major inter-city connections. Whilst we could have used a year-by-year analysis, we felt this was over analysing the problem as collaborations are built up over a long time period and synchronicity with a particular year is unnecessary. The data consists of flight volumes between airports, with 9192 airports and 33075 flight links between them for the year that we focus on. 

\subsubsection{Comparisons}
\begin{figure}
\centering
\subfloat{\includegraphics[width=\textwidth]{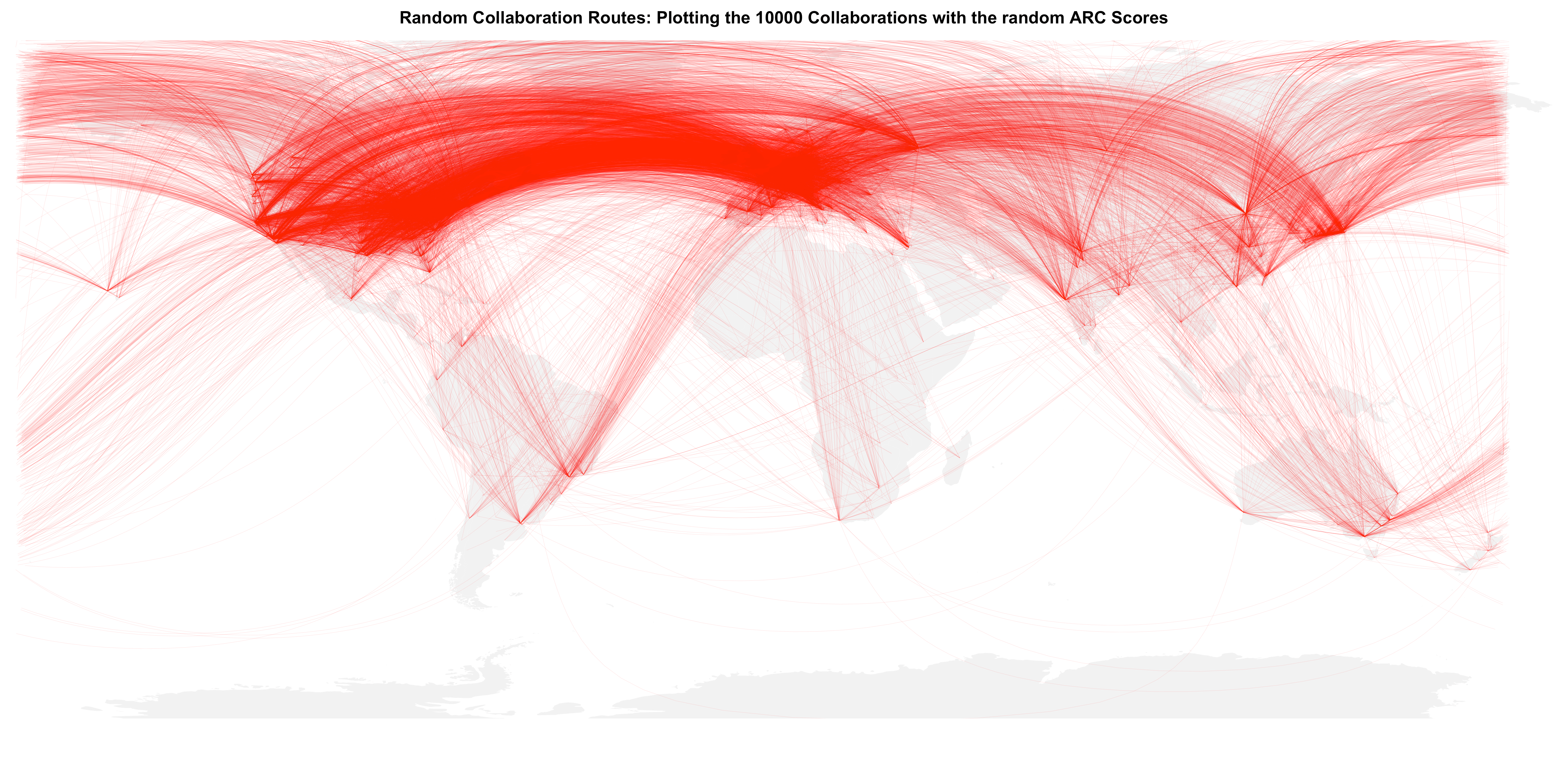}}
\caption{Global collaboration route plots}
\label{fig:global_patterns}  

\end{figure}

\begin{figure}
\centering
\subfloat{\includegraphics[width=\textwidth]{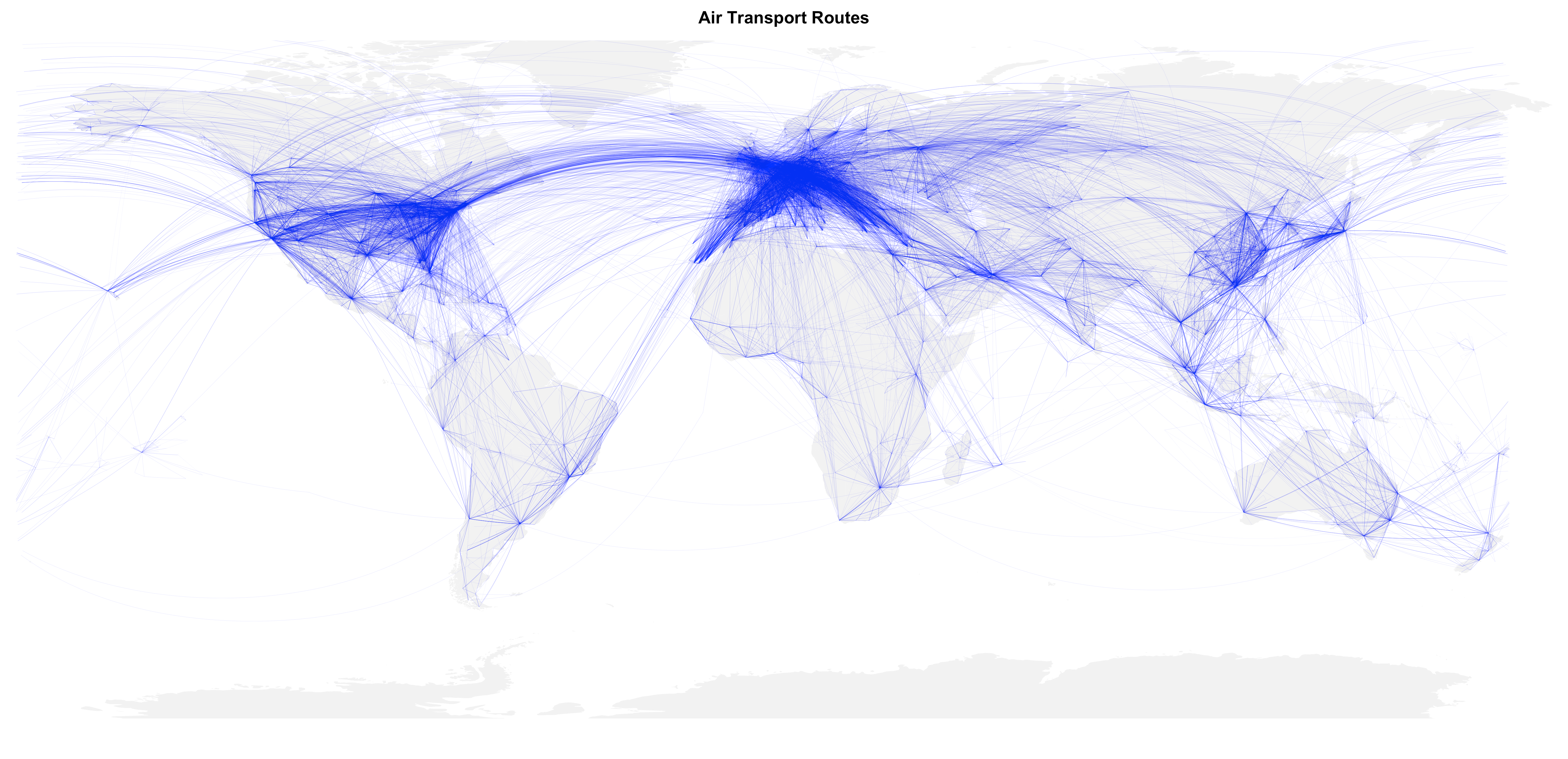}}
\caption{Global air transport route plots}
\label{fig:global_flights}  

\end{figure}

In Figure~\ref{fig:1} we see some simple comparisons between the networks of interest. We explore some of these in more detail here. In Figure \ref{fig:global_patterns} we see a random sample of the collaboration routes (the total number of routes is too large to plot clearly), whilst in Figure \ref{fig:global_flights} we see the air transport routes. Comparing these, we see a number of differences. Firstly, we see that although there is a strong connection between the US and Europe in the air transport network, this is far more pronounced in the collaboration network. The same pattern holds true for the connections between Europe and Asia and Asia and the US. Indeed, if we restrict ourselves to collaborations with coauthors from two or three different cities, we can see from Table \ref{table:top_paper_collaborations} that the top collaboration routes (by ARC score) follow these patterns. 

As noted in Figure~\ref{fig:1}, we see a north-south divide in the data, with disproportionately many collaborations occurring between cities in the global north. In particular, the percentages given in Figure~\ref{fig:1}(b) are calculated by considering every pairwise collaboration, and noting the location of the two relevant collaborators.

\begin{table}[htb]                                                             
\centering                                                         
\begin{tabular}{|l|c|l|c|}
  \hline
 \multicolumn{2}{|c|}{Two way collaborations} &  \multicolumn{2}{|c|}{Three way collaborations}  \\ 
  \hline
Countries & No. of collaborations & Countries & No. of collaborations \\
\hline
Canada-USA	&3447&	Germany-UK-USA	&128\\
Germany-USA&	3043	&France-Germany-USA&	108\\
UK-USA	&2965	&Germany-Switzerland-USA	&106\\
China-USA&	2578&	Canada-UK-USA&	93\\
Japan-USA&	2252&	France-UK-USA	&93\\
   \hline
\end{tabular}

\caption{Top two and three way collaborations by country}                                                         
\label{table:top_paper_collaborations}                                                       
\end{table}  

From this preliminary analysis, we also notice that there are a lot of long-distance collaborations present, in many cases between cities that do not have direct flights between them. This raises the interesting question of how journeys with multiple flights act as a barrier to collaboration, and what role is played by the distance on the air transport network, compared with Euclidean distance. This provides further motivation for our work.


When performing our full analysis, our focus is on linking the number of citations that each paper receives with the relationship between the coauthors on the air transport network. More specifically, we want to see if there is a link between some measure of geographical diversity of the coauthors via the air transport network, and the ARC score for the paper. Thus, in what follows we split our data by paper rather than considering summaries over all papers collaborated on by pairs of cities. For each paper we then have access to a list of the coauthors on it, their location, and the ARC score. This is what we use for our analysis.

\subsubsection{University Rankings}
One more dataset that we will make use of is the world university rankings, which comprises the rankings of the top 500 universities each year from 2005 onwards. As before, we focus on data from the year 2005. This data is necessary for our analysis because, as shown by \cite{clauset2015systematic}, there is a relationship between the reputation and ranking of a university and the number of citations that a paper written by one of its researchers receives. When we look for a relationship between the number of citations that a paper receives and our various measures of diversity of the coauthors, we want to make sure that we take this effect into account.

\subsection{Analysis} \label{analysis}
We now present the methods we use to investigate the link between geographical diversity of coauthors on a paper and the number of citations it receives. A key part in this will be defining our measures of geographical diversity. The first step towards these definitions is to connect our coauthorship data with our air transport data. 

\subsubsection{Connecting Cities with Airports}
There are a number of different ways to connect the coauthorship data with the air transport data. Firstly, we want to find a distance measure between the cities in the coauthorship dataset, where this distance is linked to the air transport network. We do this in an effort to replicate how two collaborating authors from potentially different countries could travel in order to meet each other. An initial measure of the distance between two cities is the number of flights it takes to travel between the two. We can calculate this by mapping each city to an airport and then finding the graph distance between the two airports on the air transport network. 
 
We can improve upon this by incorporating Euclidean distances between the nodes of a graph, as in \cite{gastner2006spatial}. This is done by assigning an effective length to each edge
\begin{align} \label{eq:euc_weights_app}
\text{effective length of edge } (i,j) = \lambda d_{ij}+(1-\lambda) 
\end{align}
where $d_{ij}$ is the Euclidean distance between nodes $i$ and $j$, and $\lambda$ is a parameter that controls the relative importance of physical distance against graph distance. The weighted network distance between two nodes is then given by the sum of the effective lengths on the shortest effective path between them. Incorporating Euclidean distance into our model makes sense intuitively because our distance measure is attempting to capture geographical diversity of coauthors. We believe an important part of this is the difficulty of two potential collaborators traveling to meet each other. With this in mind, a long haul flight presents more of a barrier than a shorter one. 

It can be shown that, for the global air transport network, the value of $\lambda$ that leads to the best replication of the observed network is $0$ or close to it \cite{gastner2006spatial}. In our model, we choose $\lambda=\frac{1}{10000}$. This choice fits with the conclusions of \cite{gastner2006spatial}, but is also useful from a practical perspective. We measure the Euclidean distances in kilometers, and since the longest distance Euclidean distance between two nodes on the air transport network is $\sim 9000\text{km}$ this means that a journey that involves multiple flights will always be assigned a greater weighted network distance than one involving only a single flight. Again, this fits with our intuition about the difficulty of two potential collaborators meeting, and gives some interpretability to the weighted network distances.
 
Using this, we calculate the weighted network distance between two cities $A$ and $B$ suing the air transport network as follows:
\begin{enumerate}
\item \textbf{Mapping Cities to Airports} - First, each city is mapped to one or more airports chosen as follows. We calculate the weighted degrees, on the air transport network, of all the airports within $100$km of the city. The city is then mapped to the five airports with the highest weighted degrees. If there is no airport within $100$km of the city then it is mapped to the nearest airport. We denote the sets of airports associated with cities $A$ and $B$ as $\mathcal{A}$ and $\mathcal{B}$ respectively.
\item \textbf{Calculating Weighted Network Distances} - For each pair of airports $(a,b)_{a \in \mathcal{A},b \in \mathcal{B}}$ we then calculate the weighted graph distance on the air transport network using the edge weighting given by (\ref{eq:euc_weights_app})
\item \textbf{Calculate Shortest Route} - We set the weighted network distance between $A$ and $B$, which we denote as $d_{AB}$, to be the minimum of these weighted network distances.
\item \textbf{Correcting Zero Distances} - Sometimes, due to the geographical proximity of two cities, the same airport might appear in $\mathcal{A}$ and $\mathcal{B}$. In this case, the minimum calculated In Step 3 will be $0$, even though the cities may be up to $200$km apart. To correct for this, the distance between the two cities is set to be proportional to the Euclidean distance between them, normalized so that the maximum value it can take is $1$. 
\end{enumerate}
The weighted network distance between the cities $A$ and $B$ is thus defined as 
\begin{align} \label{eq:weighted_network_distance}
d_{AB}=\min_{a \in \mathcal{A},b \in \mathcal{B}} \sum_{n=1}^N\lambda d^{e}_{i_ni_{n+1}}+(1-\lambda) + d^{e}_{AB}\mathds{1}_{\mathcal{A}\cap \mathcal{B}\neq \emptyset}
\end{align}
where $d^e_{ij}$ is the Euclidean distance between $i$ and $j$, and $a=i_1 \to i_2\to \ldots \to i_N=b$ is the shortest weighted path from $a$ to $b$ on the air transport network.
 
We choose to map each city to potentially multiple airports in another attempt to recreate real world travel situations, since the nearest airport to a city may not be the one with the best connections to certain other cities. The $100$km limit is set as the limit that a person might be willing to travel to an airport. Using a similar intuition to our choice of $\lambda$, setting the maximum distance to be $1$ in the case that two cities share an airport is to ensure that any journey that contains a flight is considered `longer' than one that does not. 

In Table \ref{table:corr}, we can see that the weighted airport network distance is quite highly correlated with the Euclidean distance. When comparing ARC scores with Average Distance for different values of $\lambda$, we will see similar patterns for varying $\lambda$. This is perhaps unsurprising given these high correlation values.
    
\begin{table}[htb]   
\centering  
\begin{tabular}{|c|c|c|c|}   
\hline                
 & Airport Network & Weighted Airport Network & Euclidean \\
\hline                
 Airport Network & 1 & 0.96 & 0.62 \\
\hline                            
 Weighted Airport Network & 0.96 & 1 & 0.80 \\
\hline     
 Euclidean& 0.62 & 0.80 & 1 \\                                  
\hline           
\end{tabular}         
\caption{Correlations between distance measures}                                
\label{table:corr}                                                        
\end{table} 

As well as using the air transport network to calculate distances between coauthors, we can use it to define centrality measures for them. Following \cite{guo2017global}, we want to find a measure of connectivity for the cities in the coauthorship dataset by associating them with airports in the air transport data set. That is, we want to find out how connected the cities are within the air transport network, as opposed to within the coauthorship network. We do this using the same method of calculating a weighted aggregate of the connectivities of each of the airports associated with a city. For any particular centrality measure $i$, such as eigenvector centrality or betweenness, the weighted centrality of a city $A$ is thus given by
\begin{align}\label{eq:weighted_centrality}
C_i(A)=\sum_{a \in \mathcal{A}}C_i(a)\left(d^e_{aA}\right) ^{-\alpha}
\end{align}
where $\mathcal{A}$ is the set of airports within $100$km of $A$, as before. $C_i(a)$ is the centrality of airport $a$, $d^e_{aA}$ is the Euclidean distance between the city $A$ and airport $a$, and $\alpha$ is a decay parameter that we set to be equal to $2$ as in \cite{guo2017global}.

\subsubsection{Connecting Cities with Universities}
As noted previously, the reputation of a university can have a large effect on the number of citations a paper written by one of its researchers receives  \cite{clauset2015systematic}. Thus, we may want to control for university rankings in our analysis. We can use the university rankings dataset to do this, but since the nodes in the coauthorship network are cities rather than universities we will have to use a similar method as we have done for the centrality measures in order to associate the ranked universities with the cities. 
 
We can construct a university rank weight for each city $A$ as follows. Firstly, we find all the universities within $20$km of the city and call this set $\mathcal{U}_A$. Then we calculate the weight $w_A$ as follows 
\begin{align} \label{eq:uni_rank_scores}
w_A=\sum_{u \in \mathcal{U}_A}1+ \frac{1}{\sqrt[]{r_u}}
\end{align}
where $r_u$ is the rank of the university $u$. 
  
There are a number of things to note about this construction. Firstly, we do not use a decay factor. This is because we are trying to replicate how the coauthorship data is aggregated into cities. Here, the collaborations from a city are the collection of the collaborations from each university associated with that city, with no dependence on how far the universities are from the city. Since we do not know exactly which universities are associated with each city, we use $20$km as an estimate. Empirically, this seems to include the relevant ranked universities for the largest cities of interest. The downside of this method is that many small towns very close to much larger cities are also given high university rank weights. This is hard to avoid with the current method, since all we have to match cities with universities are the respective location coordinates. Moreover, this will not affect our results significantly because these smaller towns have relatively few edges in the coauthorship network, except in the case when they are home to a large university. In this case, the large university ranking weight will have been assigned to them correctly. 

The exact form of the weight with respect to the rankings is calculated so that the better a ranking is, the more weight it adds, with the square root term ensuring that this effect is not too dominant. We only have the rankings for 500 universities, and so for most cities the university set $\mathcal{U}_A$ will be empty. The $+1$ means that the baseline weight is $1$ rather than $0$, since for a specific paper we may want to look at the product of the university rank weights for its coauthors. For example, a city which did not have any top 500 universities within its radius would have a weight of 1. Boston has the highest weight of 2.84, which is unsurprising given its proximity to Harvard and M.I.T.
\subsubsection{Measures of Diversity}
We now present the three measures that we will use to investigate the relationship between coauthor diversity and paper citations.
\begin{enumerate}[label=(\alph*)]
    \item \textbf{Average Weighted Network Distance:} We have already outlined a method for calculating a weighted network distance between two cities. For a specific paper $P_i$ with $N_i$ coauthors from cities $c_{i1},\ldots,c_{iN_i} \in \mathcal{C}_i$ we can then calculate the average weighted network distance as
    \begin{align}
    \frac{1}{\binom{|\mathcal{C}_i|}{2}} \sum_{c_{ij},c_{ik}\in \mathcal{C}_i}d_{c_{ij}c_{ik}}
    \end{align}
    which is the average of the weighted network distances between all the pairs of coauthors on the paper. This is a simple measure, but it captures the geographical diversity of the coauthors in a sense which takes into account the difficulty of traveling between their various locations. The intuition behind it is also clear - a higher average weighted network distance means that on average the coauthors are further apart both geographically and in terms of travel links, and are thus more diverse in this sense.
    \item \textbf{Entropy of Weighted Network Distance:}
    A related measure of diversity can be found by calculating the entropy of the weighted network distances between the coauthors on a paper. We use the Shannon entropy \cite{shannon1948mathematical}, defined as 
    \begin{align}
    H= -\sum_i p_i \log(p_i)
    \end{align}
    where the $p_i$ in this case are the probability of a certain weighted network distance appearing given the distribution of distances in our data. We can estimate these probabilities by sorting the observed distances into bins and then using the bin counts as an empirical distribution estimator. 
     
    This measure, also known as Shannon's diversity index, quantifies the diversity of weighted network distances between coauthors on a paper. It may be more difficult to see how this measure captures diversity in a similar sense to our previous measure. In this case a larger value indicates that the distances between coauthors are more varied. From the viewpoint of one specific coauthor, this would indicate that they collaborate with coauthors that are varying distances away from them - perhaps one international coauthor and one from a nearby university. Conversely a smaller value would indicate several coauthors that are the same distance from each other, such as several coauthors from local universities. It is worth noting that this measure is only meaningful for papers with more than two coauthors. With only two coauthors this entropy measure will always be zero, as the entropy of a single number is zero.
    \item \textbf{Weighted Entropy of Coauthor Location:} An entropy-based measure that may seem more intuitive can be found by directly calculating the entropy of the geographical locations of the coauthors of a paper. We can calculate this as before by discretising the locations into `bins' which are 2-dimensional in this case. The entropy of the locations then gives a direct measure of geographical diversity, since a higher value means that the coauthors are more spread out throughout the world, with fewer located close together in the same `bin'. This entropy measure is different to the one used previously in that it does not concern the actual (weighted network) distances between the coauthors, just whether or not they are clustered together.
     
    This initial construction does not involve the air transport network distances between coauthors or the university rank weights of their locations, both of which we have said are important factors. Thus we can improve it by using the weighted entropy introduced by \cite{guiacsu1971weighted}. This is of the form
    \begin{align}
    H= -\sum_i w_i p_i \log(p_i)
    \end{align}
    where the $p_i$ are the probabilities of a certain geographic location bin. The $w_i$ are weights which in our case take the form
    \begin{align}
    w_i=\frac{C_{eig,i}^{0.05}}{U_i}
    \end{align}
    Here, the $U_i$ are the averages of the university rank weights of the coauthor locations in the 2D bin used to calculate $p_i$. The $C_{eig,i}$ are averages of the eigenvector centralities over the bins. We `power down' $C_{eig}$ by raising it to a small power because the range is huge (over 10 orders of magnitude) and we do not want it to dominate the entropy values or university rank weights.
      
    This form for the weights associates more weight with lower ranked universities and less connected cities. Thus, our measure of diversity rewards papers where the coauthors are not only spread out geographically but also not well connected on the air transport network. This means that papers with a higher weighted diversity indicates a greater difficulty for their coauthors to travel to each other, which is in line with our previous measures. The diversity measure also rewards papers with coauthors from less highly ranked universities, which helps to counteract the effect reported by  \cite{clauset2015systematic} on the effects of university rankings and reputation on citations. 
      
\end{enumerate}

\section{Discussion on Impact}\label{sec:impact}
In terms of impact on the academic knowledge transfer and international collaboration, there are two distinct areas that these results can contribute to. 
(1) Exchange and Mobility: many bilateral schemes (e.g., Royal Society International Exchange, German DAAD) dictate which countries are priority countries based on largely bi-lateral funding agreements and a common scientific priority agenda. Often this overlooks diversity and especially the global north-south divide highlighted in this paper (94\% of collaborations are between northern hemisphere universities). Beyond travel grants, domain specific researchers can also benefit from this work (e.g., which countries have the greatest diversity potential for similar distance).
(2) Inform Research Funding Policy: current best practice recognises the need to improve diversity, but lacks quantitative frameworks. Whilst this work only provides a single dimensional geographic diversity (though one can argue geography is closely associated with many aspects of culture, ethnolinguistics, and practices), it provides domain specific data on diversity gaps. This in turn can inform both university policy as well as add an extra diversity dimension for international partnerships (e.g., current GCRF funding is only based on income).

\section{Conclusions} \label{conclusion}
In this paper we have investigated connections between citations that papers receive and how the coauthors are connected via the air transport network. In particular we have looked at how different measures of geographical diversity of the coauthors on a paper are related to its ARC score. We have defined three different measures of diversity, relating to the average weighted (air transport) network distance between coauthors, the entropy of these weighted network distances and the weighted entropy of the coauthors' geographical locations. We have seen interesting relationships in each case. For the two types of entropy, the average ARC score for a paper increases as the entropy, and thus the diversity, increases. As the average weighted distance increases, the ARC scores increase up to a point, but then start to decrease. In all cases there appears to be a link between diversity and citations.

To ensure that there were no obvious global confounding variables that could offer an alternative explanation for these results, we have also investigated the effects the university rankings have on this relationship. We have seen that the relationship between the diversity measures and the average university rank weights is similar to the relationship between the diversity measures and the ARC scores. However, we have shown that the effects discussed above persist having controlled for the effects of university rankings. Furthermore, we have seen that different subject areas exhibit different relationships between diversity and ARC scores. This is also true when we look at collaborations made by researchers from specific cities.

\section{Author Contribution}
C.N. carried out the analysis. V.L. and C.R.S. provided the citation data and supported the interpretation of the data and results. W.G. conceived the diversity idea and provided the air traffic and airport data. C.L. and W.G. worked with C.N. to design the data analysis. All authors contributed to writing the paper. 

\quad

\textbf{Data Availability}
The datasets used in our analysis are available from the corresponding author on reasonable request.

\quad

\textbf{Competing Interests}
The authors have no competing interests to report.

\bibliographystyle{IEEEtran}
\bibliography{IEEEabrv, bibfile}

\appendix

\section{Additional Results}\label{additional_results}
\subsection{Robustness to choice of $\lambda$}

When constructing our measures of diversity, the most important parameter that we choose is $\lambda$, which controls the balance between Euclidean distance and flight hop distance in Equation (3) in the main report. We know from prior evidence that $\lambda$ is sufficient to describe transport networks using simulated annealing \cite{gastner2006spatial}. In our case, we choose a value of $\lambda=\frac{1}{10000}$, as this gives some interpretability. However, we want to make sure that our results are robust to the choice of $\lambda$. To check the robustness, we repeat our results for a range of values of $\lambda$, and note whether the same relationships appear. 
\begin{figure}[t]
\centering
\subfloat[$\lambda=\frac{1}{5000}$ \label{fig:lambda_1_5000}]{\includegraphics[width=0.4\textwidth]{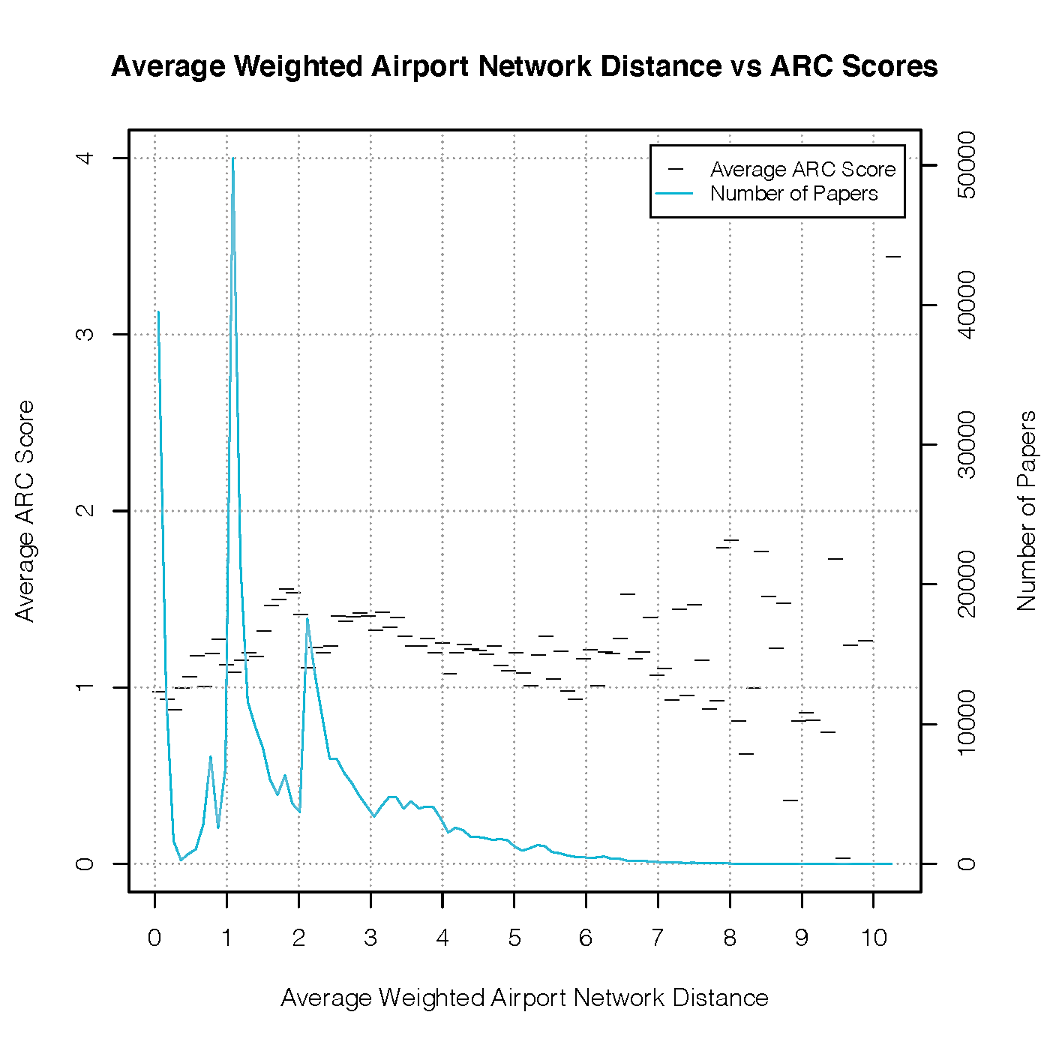}}
\subfloat[$\lambda=\frac{1}{3500}$ \label{fig:lambda_1_3500}]{\includegraphics[width=0.4\textwidth]{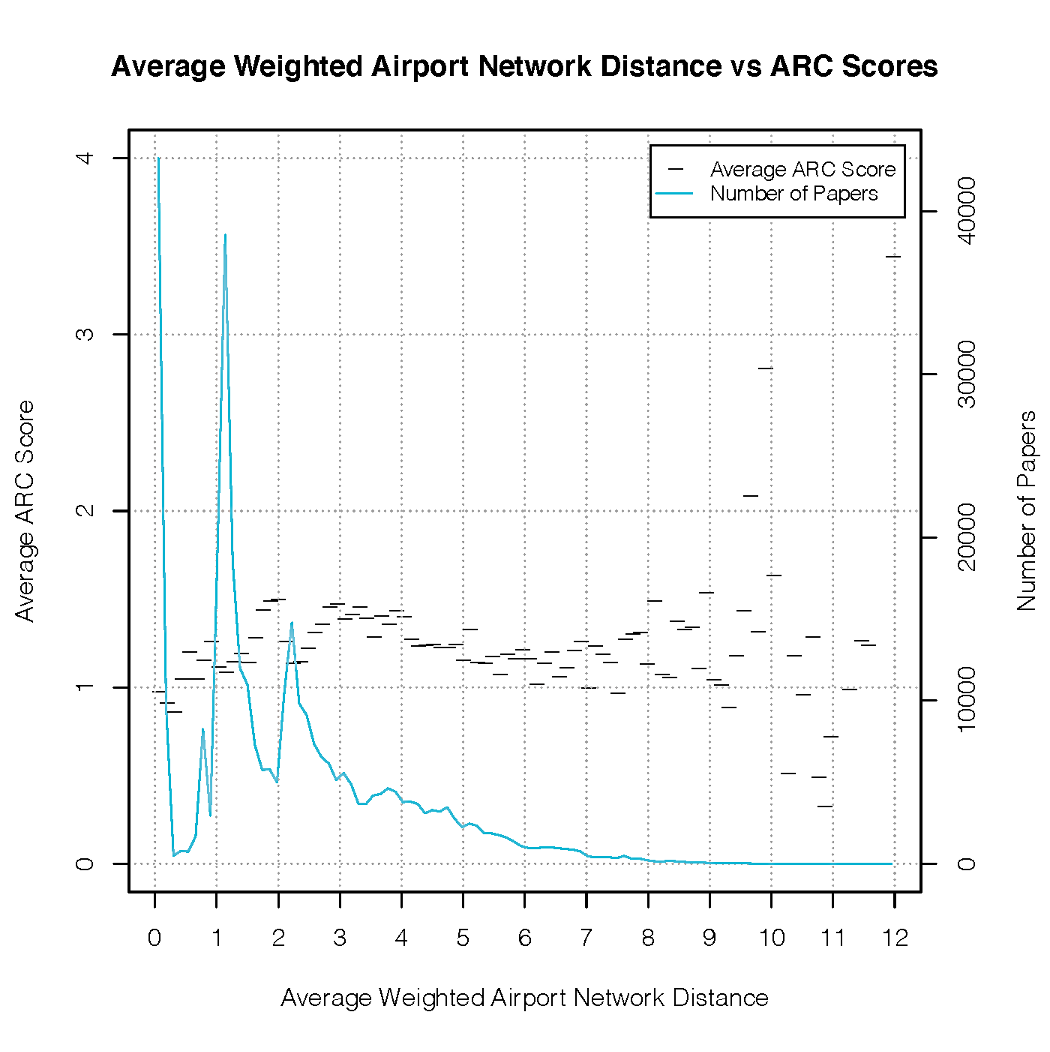}}\\
\subfloat[$\lambda=\frac{1}{2500}$\label{fig:lambda_1_2500}]{\includegraphics[width=0.4\textwidth]{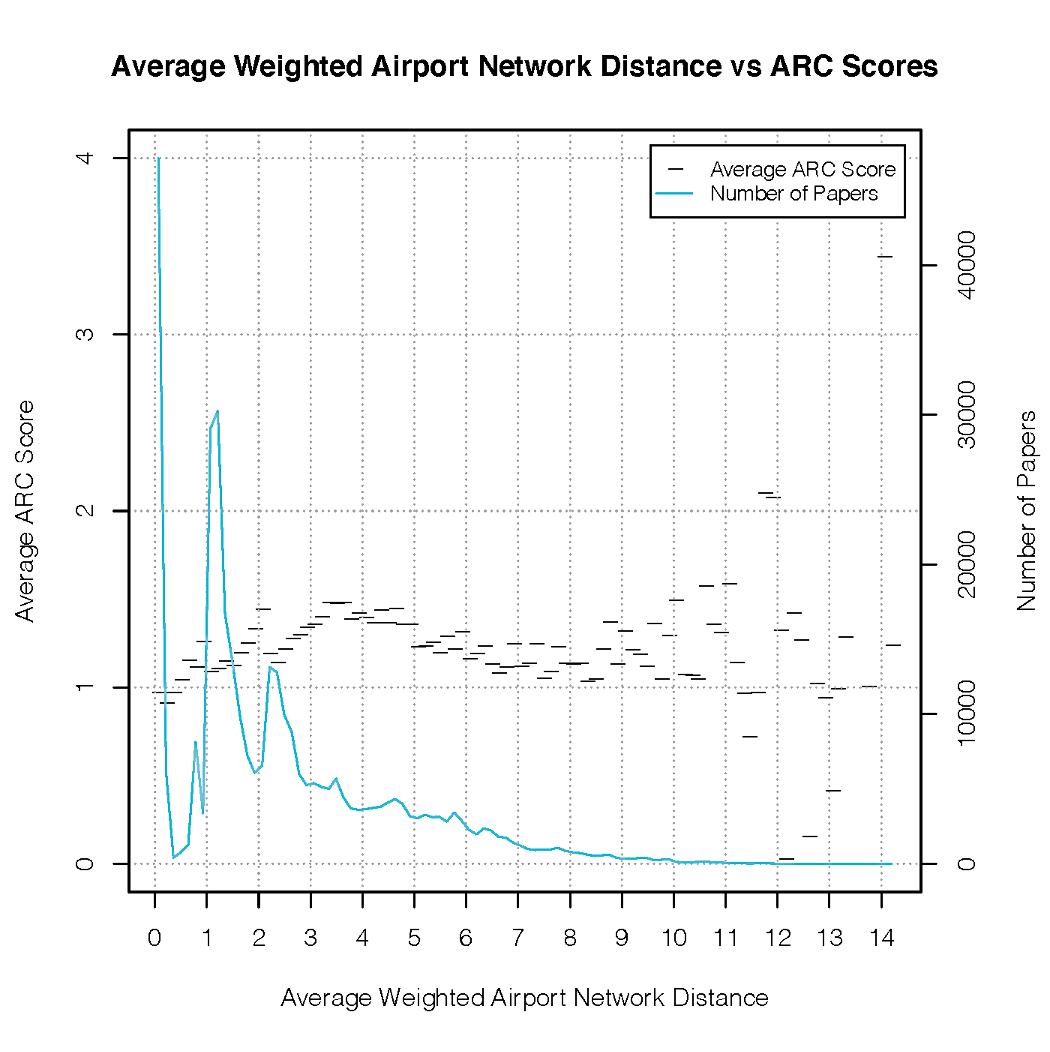}}
\subfloat[$\lambda=\frac{1}{5}$\label{fig:lambda_1_5}]{\includegraphics[width=0.4\textwidth]{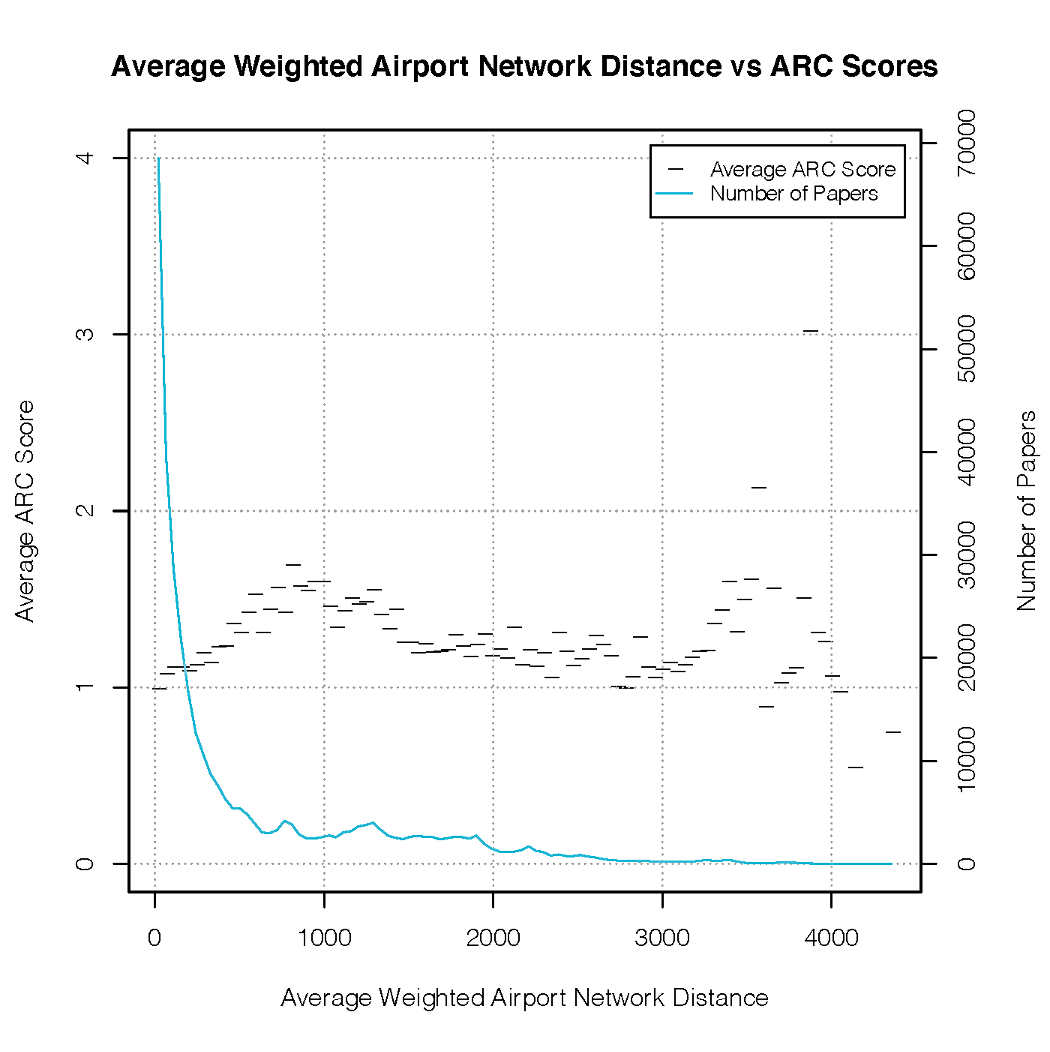}}
\caption{Relationship between average weighted airport network distance and average ARC score for $(a)$ $\lambda=\frac{1}{5000}$, $(b)$ $\lambda=\frac{1}{3500}$, $(c)$ $\lambda=\frac{1}{2500}$, and $(d)$ $\lambda=\frac{1}{5}$.}  
\label{robustness}  
\end{figure}
In Figure \ref{robustness}, we check the robustness of our method against changes in $\lambda$. We see the same pattern emerging in each case, with only the location of the peak changing as the weighted network distances are calculated differently. As mentioned previously, our choice of $\lambda=\frac{1}{10000}$ gives some interpretability which we lose for larger choices. In particular, for $\lambda=\frac{1}{5}$ the weighted distances are completely dominated by the Euclidean distances between coauthors (or more precisely, the sum of the Euclidean distances between the stops on the air transport route between them). In this case we lose the interpretation of ``Well-trodden paths'', since the effect of transport routes is greatly lessened compared to our original choice. However, it is encouraging to see a clear trend even in this case.

\begin{figure}[htb]
\centering
\includegraphics[width=0.6\textwidth]{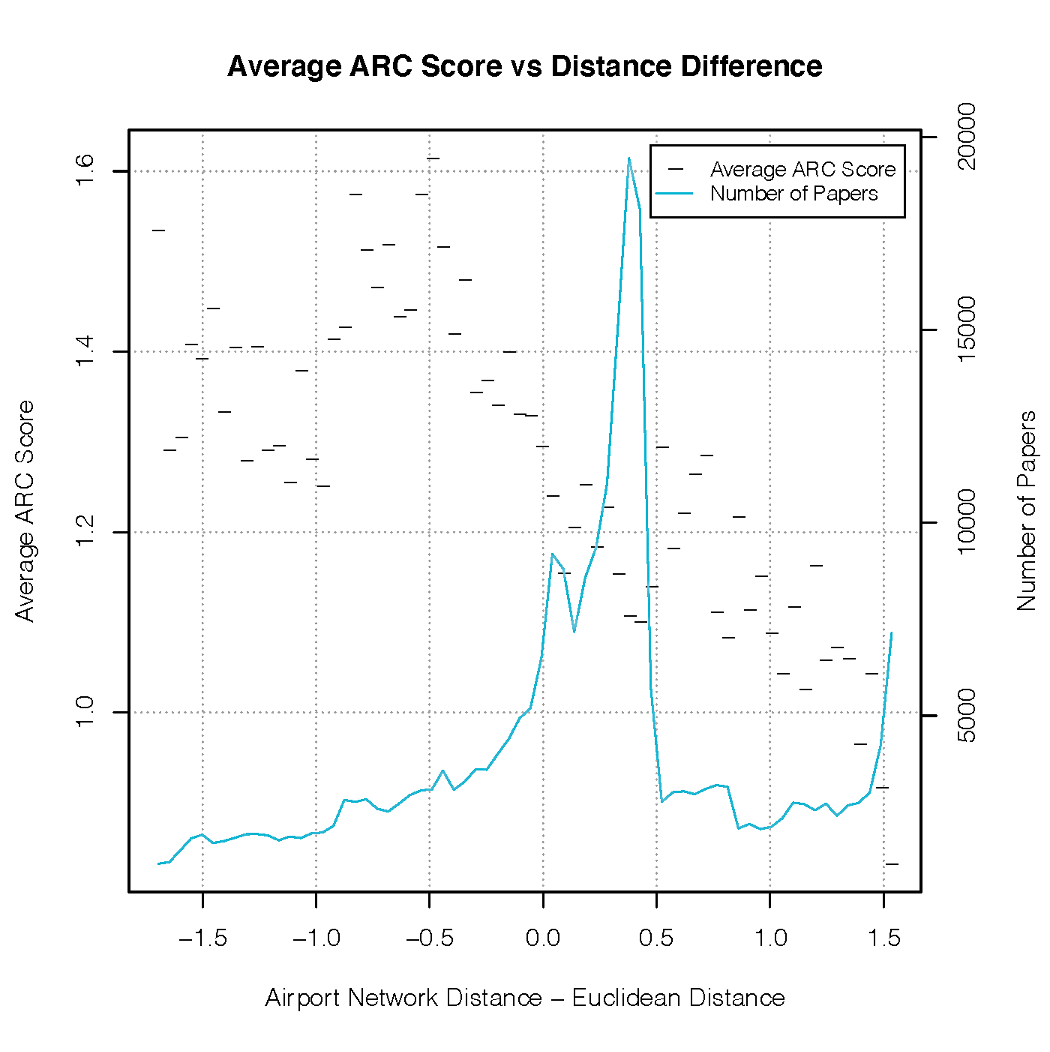}
\caption{Average ARC score vs difference between weighted network and Euclidean distance}                            
\label{fig:diff}  
\end{figure}

We may also be interested in the case where, instead of using our weighted network distance, we simply use the Euclidean distance between coauthor locations. Indeed, previous work has been done to show the positive effect of distance on collaborations \cite{nomaler2013mgore}. However, we are more concerning with identifying an effect related to the ease of travel on the air transport network. To try and disentangle these effects, we compare the weighted network distances and Euclidean distances in detail. To do this, we first map each of them to $[0,1]$. Then, having taken the difference between them, we normalize this so that it has mean 0 (and variance 1). This allows us to see when the difference is more or less than average (the difference being more or less than $0$ is not interpretable because we have already mapped each distance to the unit interval). In Figure \ref{fig:diff} we see that when the difference is less than average (i.e. when airport network distance is relatively shorter than Euclidean) the average ARC scores are higher than when it is positive. 

Overall, the average ARC when the difference is $<0$ is $1.24$, and $1.14$ for when the difference is $>0$. This indicates that there is an additional effect coming from the network distances, with a collaboration that has a higher weighted network distance (relative to the Euclidean distance) having a lower ARC on average than once with a lower relative weighted network distance. Intuitively this makes sense, as we can imagine that if two routes have the same Euclidean distance, but one is relatively better connected on the air transport network, then it is this route that is more likely to foster collaboration.

\subsection{Comparing ARC Scores and University Ranks}\label{confounding_additional_results}
\begin{figure}[t]
\centering
\subfloat[Average Weighted Distance vs University Rank Weight \label{fig:dist_vs_rank}]{\includegraphics[width=0.4\textwidth]{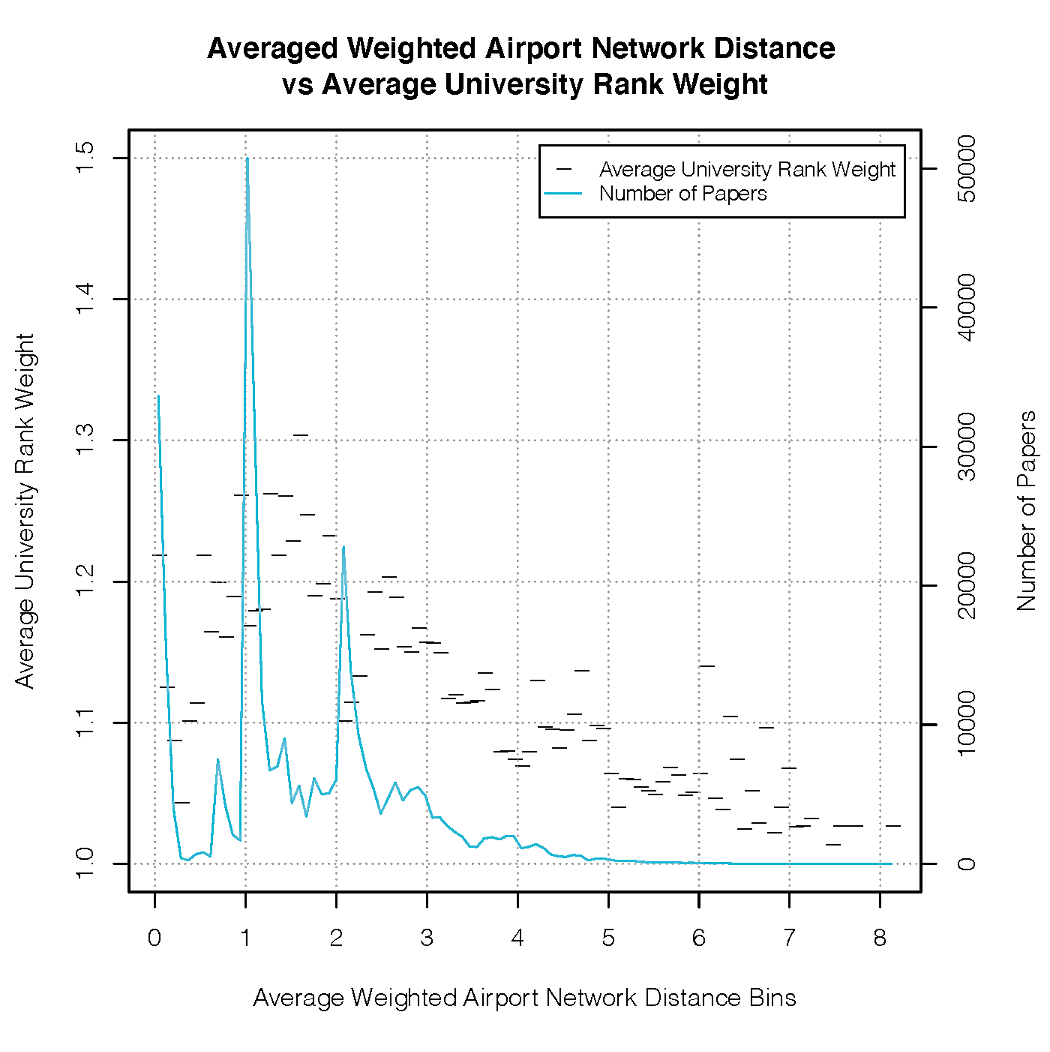}}
\subfloat[Average Weighted Distance vs ARC\label{fig:dist_vs_arc}]{\includegraphics[width=0.4\textwidth]{Fig_1a_new_indiv_bins_mean.png}}\\
\subfloat[Weighted Distance Entropy vs University Rank Weight \label{fig:dist_ent_vs_rank}]{\includegraphics[width=0.4\textwidth]{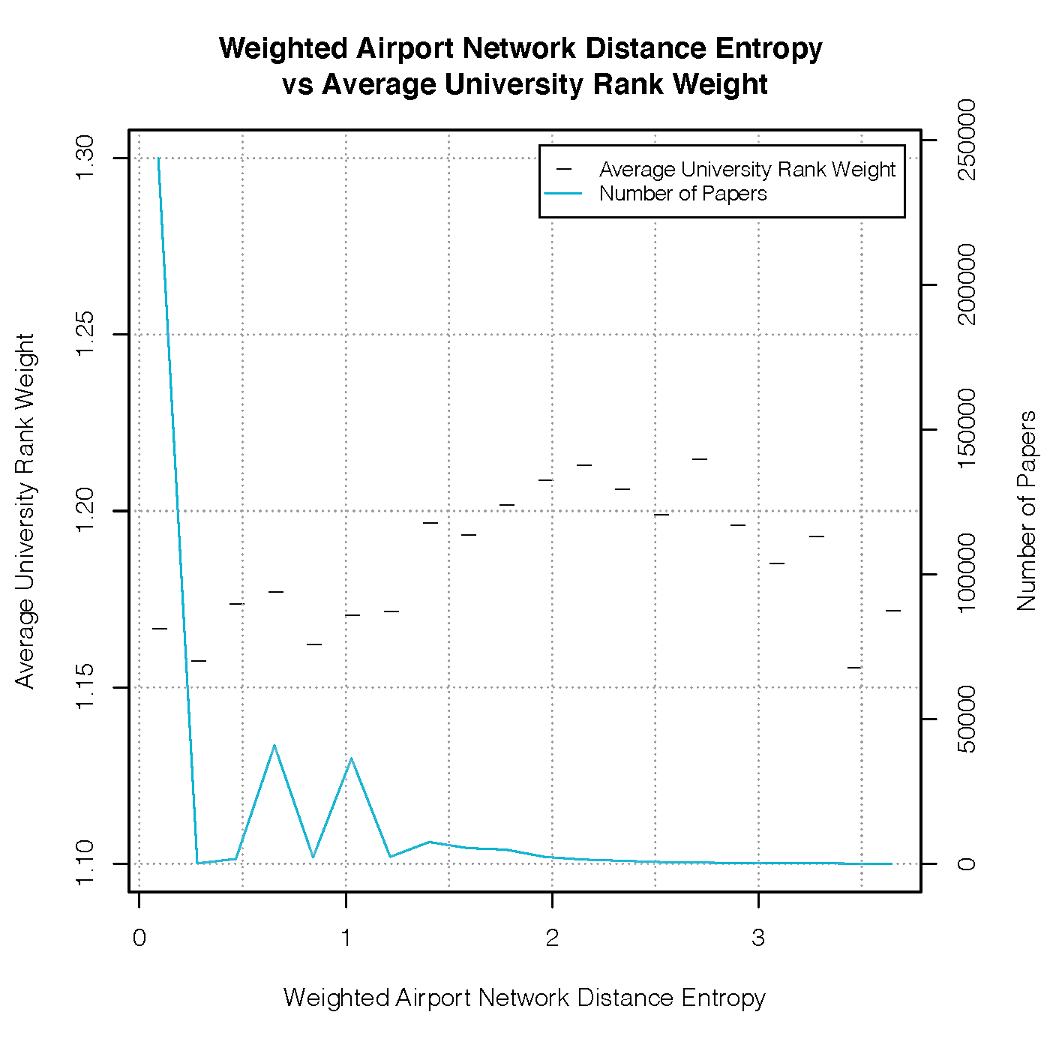}}
\subfloat[Weighted Distance Entropy vs ARC \label{fig:dist_ent_vs_arc}]{\includegraphics[width=0.4\textwidth]{Fig_1b_new_indiv_bins_ent_real.png}}\\
\subfloat[Location Entropy vs University Rank Weight \label{fig:ent_vs_rank}]{\includegraphics[width=0.4\textwidth]{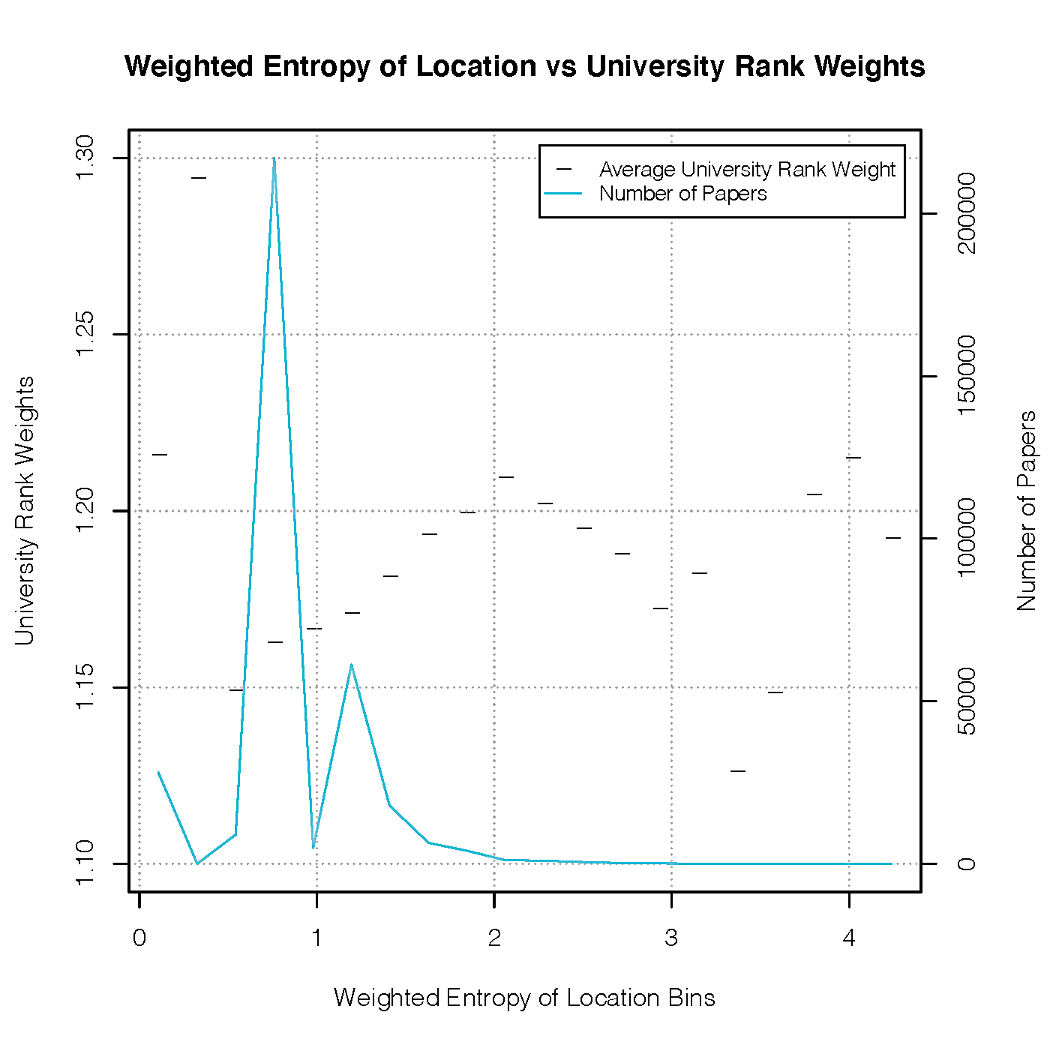}}
\subfloat[Location Entropy vs ARC \label{fig:ent_vs_arc}]{\includegraphics[width=0.4\textwidth]{Fig_1c_new_indiv_bins_weighted_ent.png}}
\caption{Comparison of diversity measures against university rank weight and against ARC score}   
\label{comparison_graphs}  

\end{figure}

In this section we perform some additional experiments to investigate the effect that university rankings have on our results. We first plot our three measures of diversity against the average university rank weights, binning as before. In Figure \ref{comparison_graphs} we compare these with the plots of the same diversity measures against average ARC score. We see immediately the similarities between them, especially for the values of each diversity measure for which the bin size is not too close to 0. We therefore see here that our various measures of diversity are having similar effects on the distribution of university rankings as they have on ARC score. This indicates the need for the analysis we perform in Section III in the main report to control for the effect of the university ranks.

Stratification is a tool that is often used to account for a confounding variable, and it is enlightening to see what happens when we stratify by the university rank weights. In Table \ref{table:strata} we see that for all strata there is a statistically significant increase in ARC Score as average weighted airport network distance increases. However, the location (or even existence) of a peak, and corresponding decrease depends quite strongly on the university ranks. In particular, we see that for lower university rank weights, there is either no peak, with both $\hat{b}_1$ and $\hat{b}_2$ estimated to be positive, or $\hat{b}_2$ is not significantly less than $0$. Conversely, for the higher strata, we do see the peaked pattern observed for the whole data set (the highest strata $(2.50,3]$ has too few data points to give an interpretable result). This suggests that while the positive effect that increasing the average weighted distance has on ARC scores is present with or without of the effect of university rankings, the subsequent decrease is at least partially tied to them. This interpretation is in keeping with what we see in Table I in the main report, with the estimates $\hat{b}_1$ and $\hat{x}^*$ remaining relatively constant, but the estimate of $\hat{b}_2$ being almost halved in value after accounting for this effect. 
\begin{table}[ht]
\centering
\begin{tabular}{|r||rrrrr|}
  \hline
 Stratum & $\hat{x}^*$ & $\hat{b}_1$ & p-value & $\hat{b}_2$  & p-value \\ 
  \hline
$(0,1]$ & 2.09 & 0.02 & 0.03 & 0.07 & 0.00 \\ 
$(1,1.02]$ & 4.22 & 0.08 & 0.00 & -0.19 & 0.50  \\ 
$(1.02,1.04]$ & 2.90 & 0.04 & 0.00 & 0.02 & 0.43  \\ 
$(1.04,1.06]$ & 1.56 & 0.12 & 0.00 & 0.05 & 0.00 \\ 
$(1.06,1.08]$ & 2.56 & 0.12 & 0.00 & 0.03 & 0.23 \\ 
$(1.08,1.10]$ & 1.85 & 0.17 & 0.00 & -0.02 & 0.55  \\ 
$(1.10,1.15]$ & 1.85 & 0.23 & 0.00 & -0.06 & 0.02 \\ 
$(1.15,1.20]$ & 1.95 & 0.36 & 0.00 & -0.16 & 0.00  \\ 
$(1.20,1.25]$ & 1.99 & 0.39 & 0.00 & -0.29 & 0.00  \\ 
$(1.25,1.50]$ & 1.68 & 0.43 & 0.00 & -0.18 & 0.00  \\ 
$(1.50,1.75]$ & 1.65 & 0.27 & 0.00 & -0.29 & 0.00  \\ 
$(1.75,2]$ & 1.70 & 0.45 & 0.00 & -0.24 & 0.00  \\ 
$(2,2.50]$ & 2.15 & 0.72 & 0.00 & -1.16 & 0.00  \\ 
$(2.50,3]$ & 0.02 & -74.58 & 0.01 & 1.18 & 0.39  \\ 
\hline
\end{tabular}
\caption{Comparison of relationships between average weighted network distance and ARC score for different strata}                                                         
\label{table:strata}
\end{table}

\subsection{Comparing Results by City}
\begin{figure}[htb]
\centering
\subfloat[Boston                                 
\label{fig:boston_piecewise_usa}]{\includegraphics[width=0.3\textwidth]{Boston_piecewise.png}}
\subfloat[Cambridge                              
\label{fig:cambridge_piecewise}]{\includegraphics[width=0.3\textwidth]{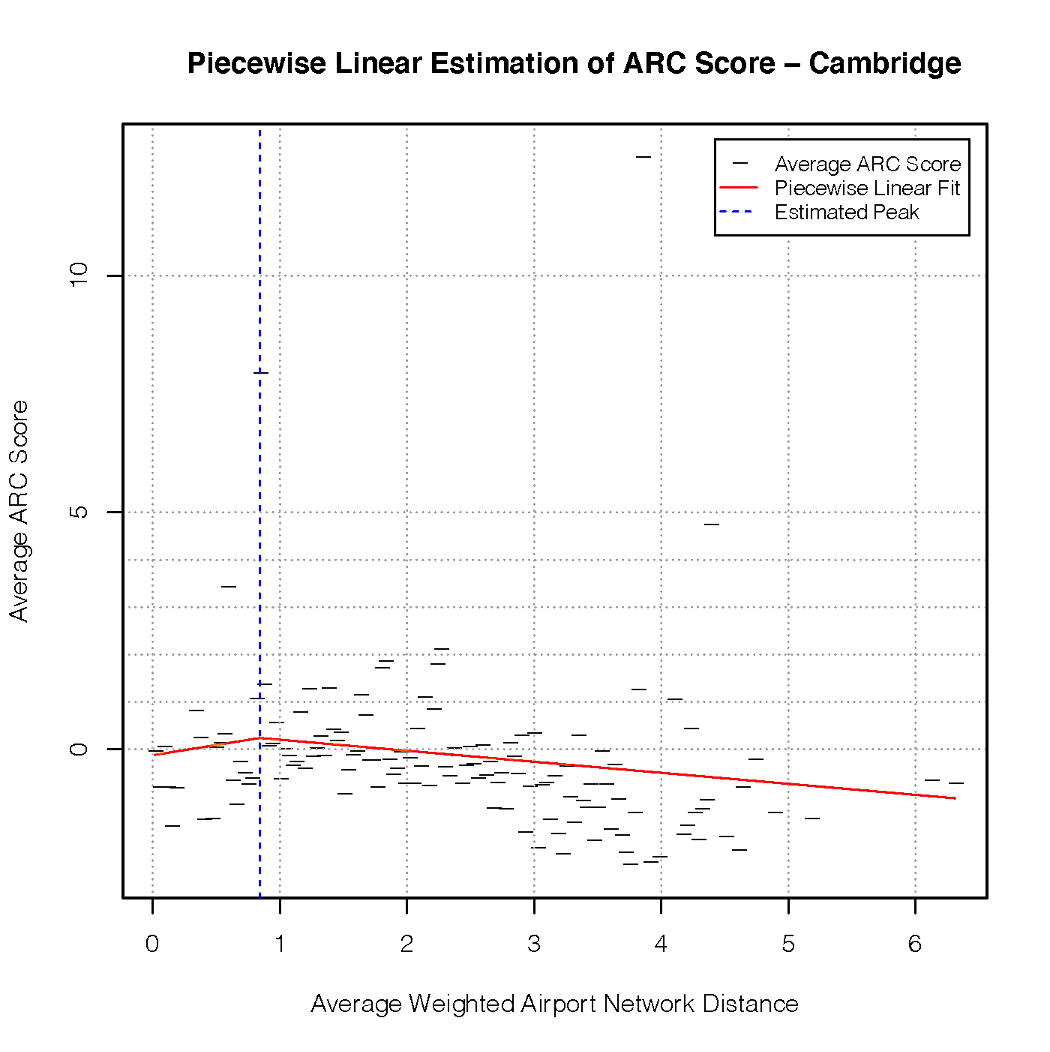}}\\
\subfloat[New York                  
\label{fig:nyc_piecewise}]{\includegraphics[width=0.3\textwidth]{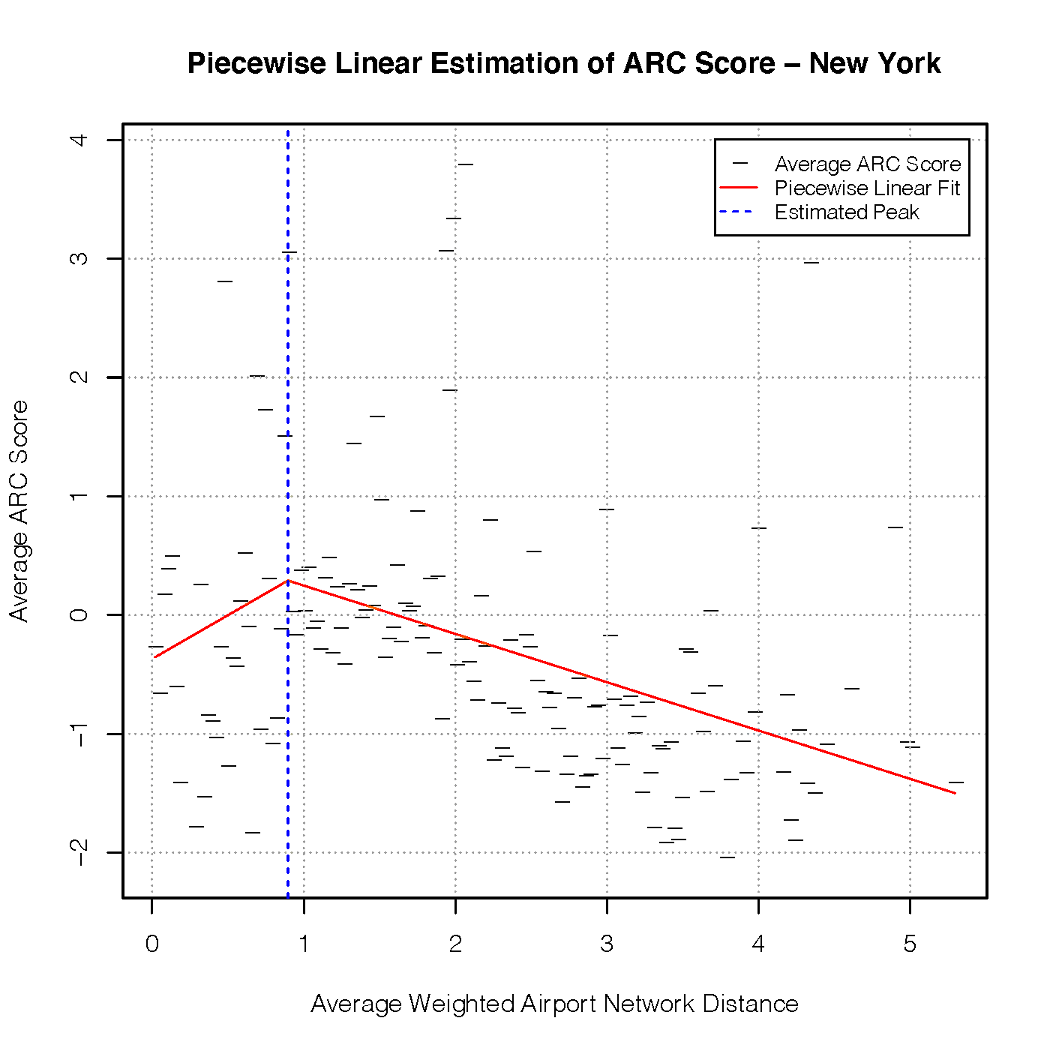}}
\subfloat[Berkeley                      
\label{fig:berkeley_piecewise}]{\includegraphics[width=0.3\textwidth]{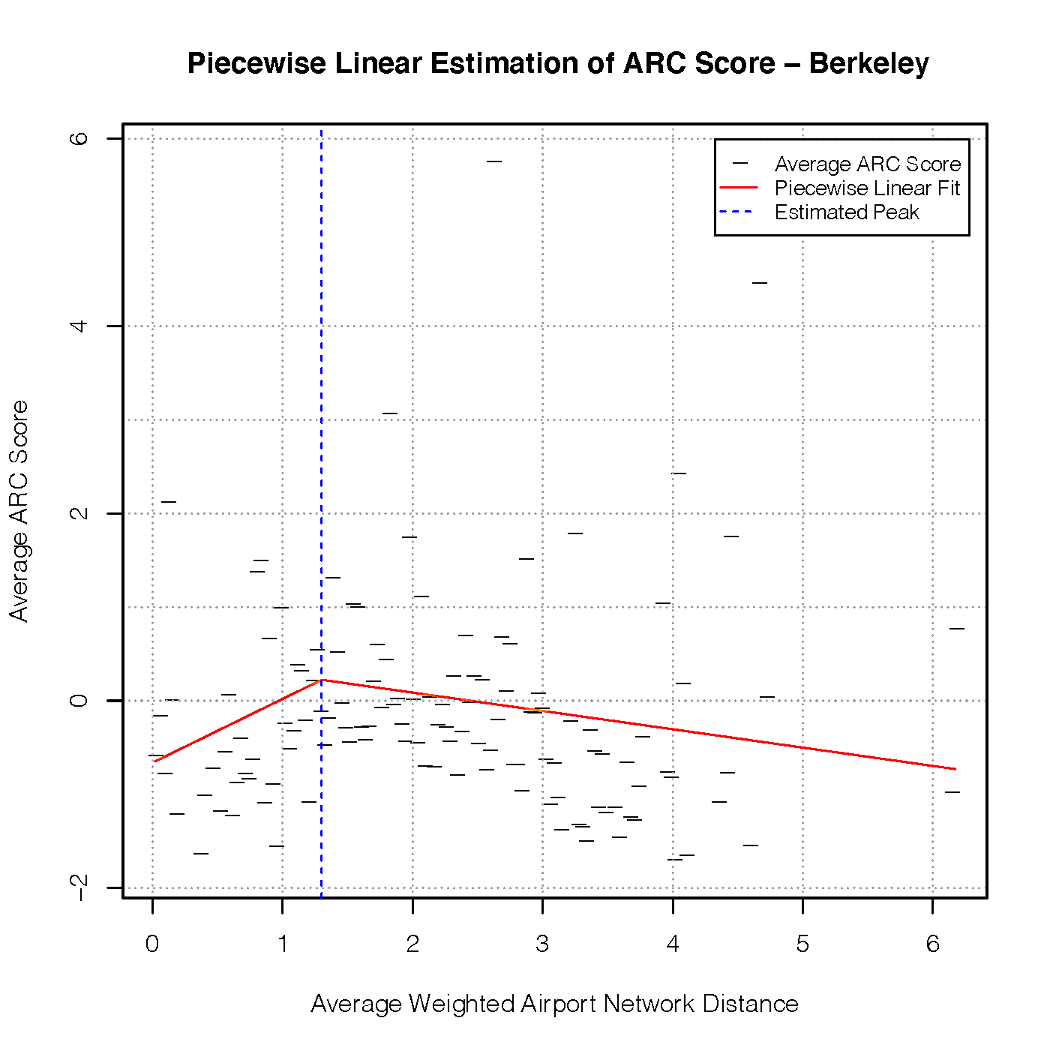}}
\caption{Piecewise linear estimation of ARC score using average weighted airport network distance, for $(a)$ Boston, $(b)$ Cambridge (USA), $(c)$ New York and $(d)$ Berkeley}             
\label{fig:usa_graphs}  
\end{figure}

\begin{figure}[htb]
\centering
\subfloat[London                                 
\label{fig:london_piecewise_uk}]{\includegraphics[width=0.3\textwidth]{London_piecewise.png}}
\subfloat[Oxford                              
\label{fig:oxford_piecewise}]{\includegraphics[width=0.3\textwidth]{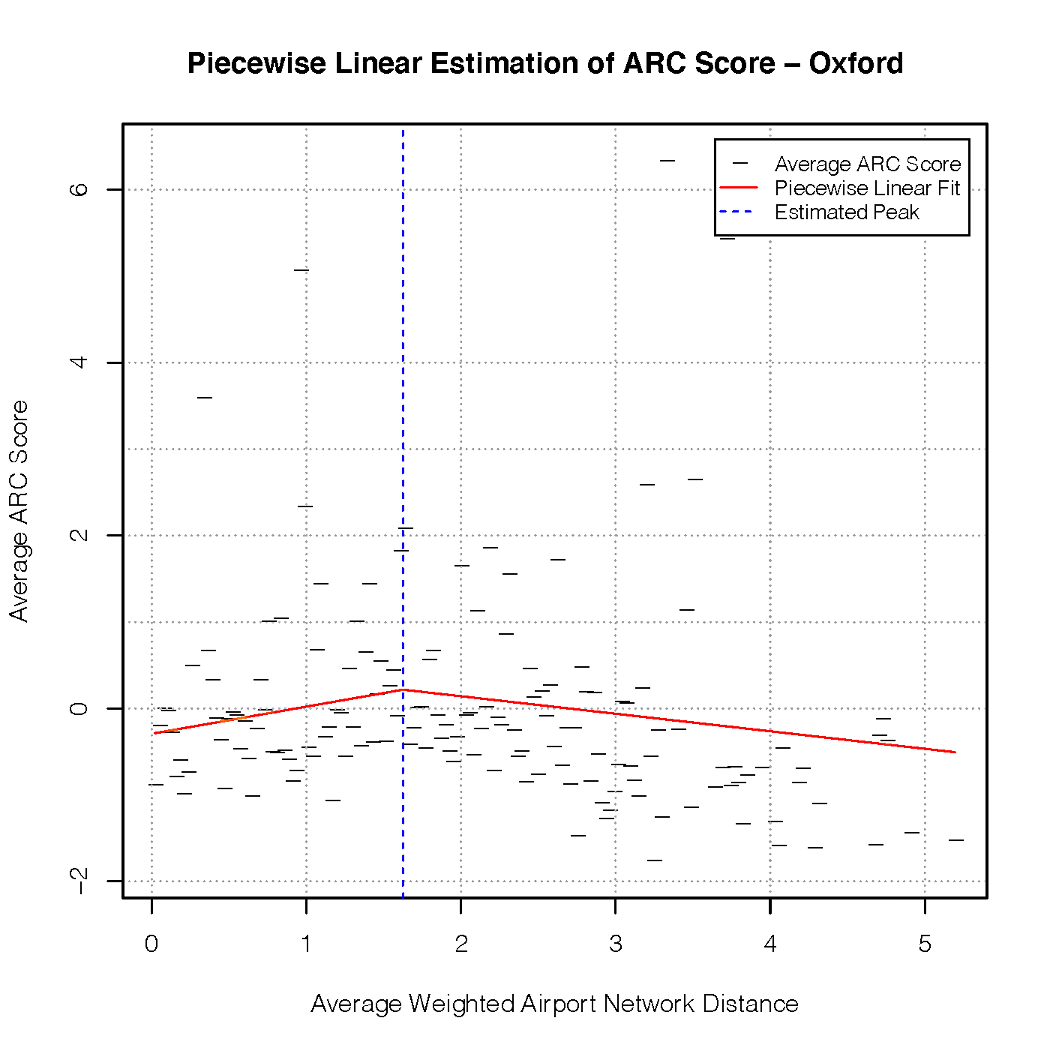}}\\
\subfloat[Edinburgh                  
\label{fig:edinburgh_piecewise}]{\includegraphics[width=0.3\textwidth]{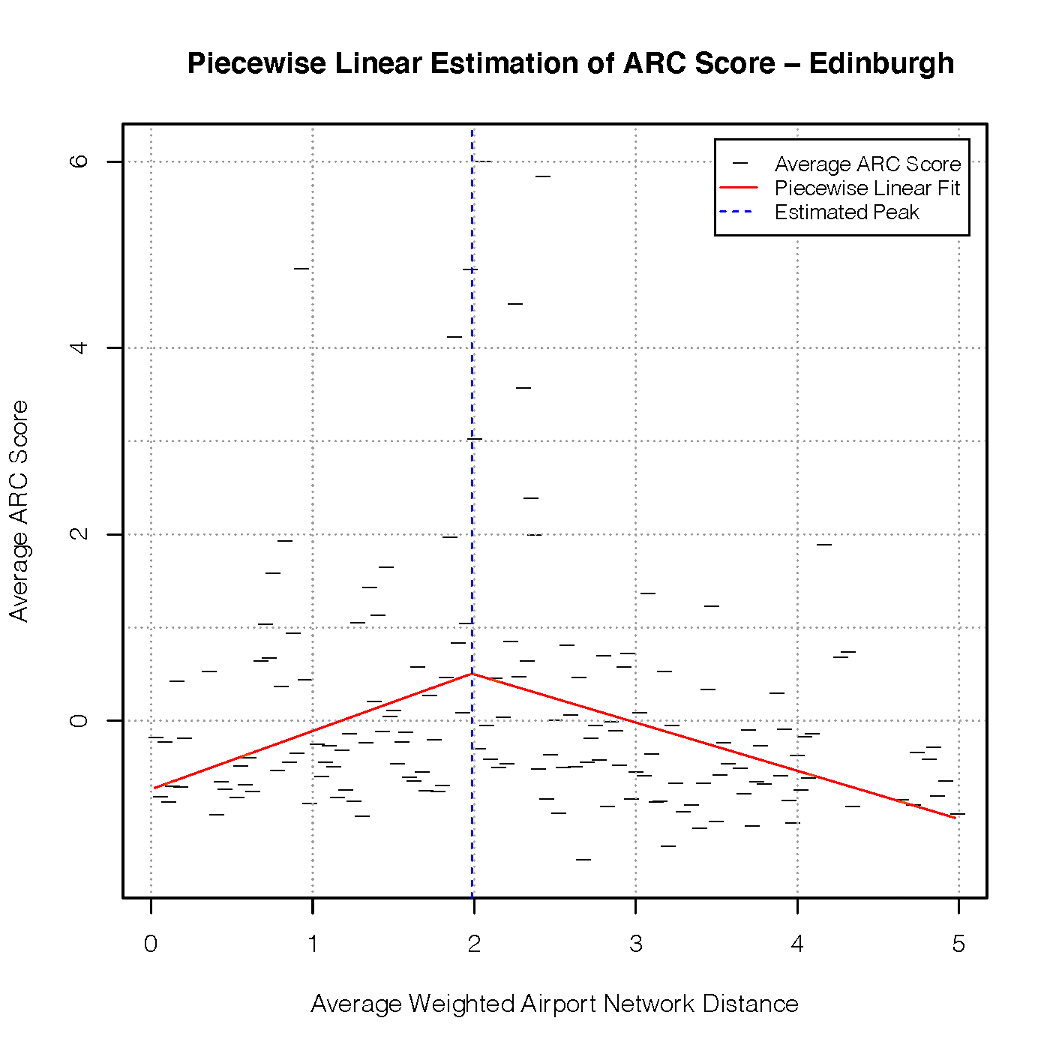}}
\subfloat[Dublin                      
\label{fig:dublin_piecewise}]{\includegraphics[width=0.3\textwidth]{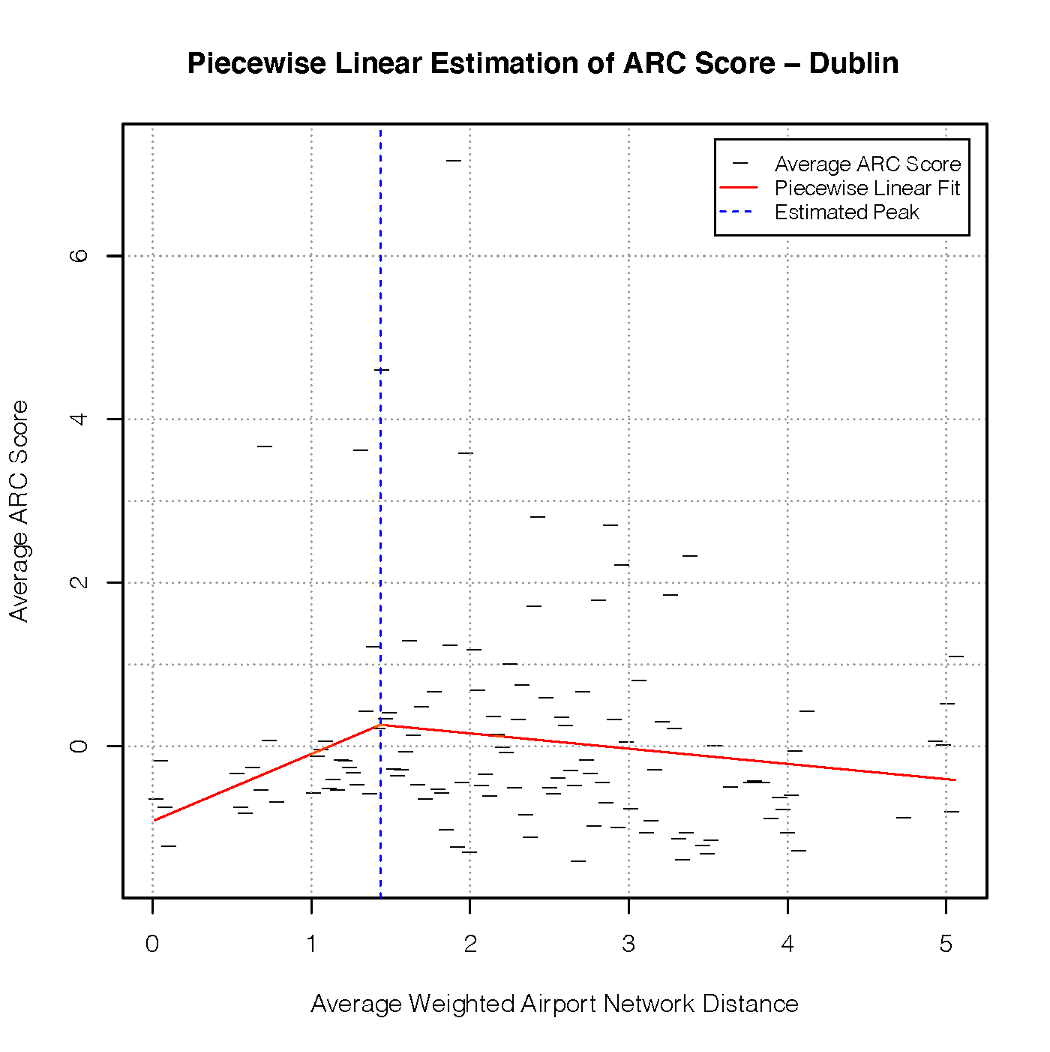}}
\caption{Piecewise linear estimation of ARC score using average weighted airport network distance, for $(a)$ London, $(b)$ Oxford, $(c)$ Edinburgh and $(d)$ Dublin}             
\label{fig:uk_graphs}  
\end{figure}

\begin{figure}[htb]
\centering
\subfloat[Beijing                                 
\label{fig:beijing_piecewise_china}]{\includegraphics[width=0.3\textwidth]{Beijing_piecewise.png}}
\subfloat[Hong Kong                    
\label{fig:hk_piecewise}]{\includegraphics[width=0.3\textwidth]{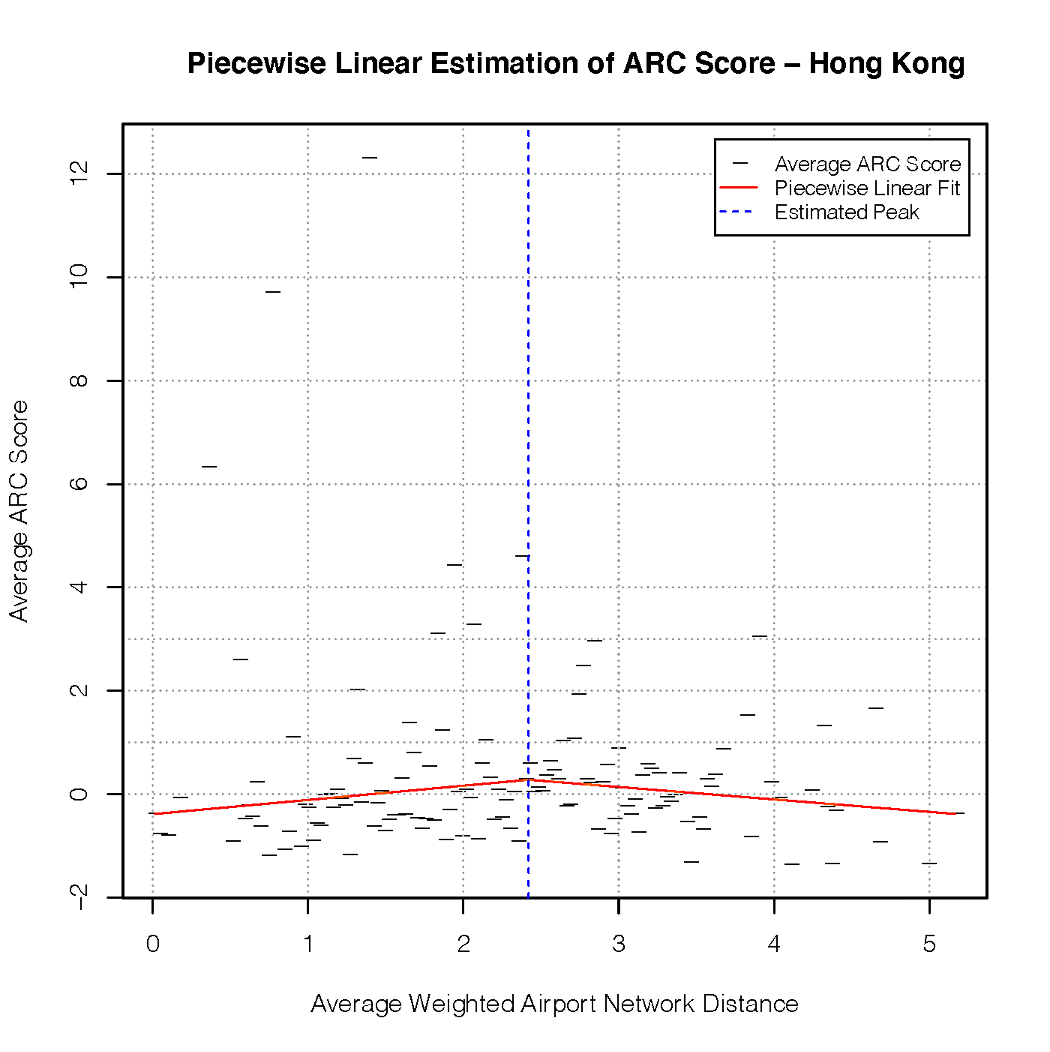}}

\caption{Piecewise linear estimation of ARC score using average weighted airport network distance, for $(a)$ Beijing and $(b)$ Hong Kong}             
\label{fig:china_graphs}  
\end{figure}

In Figures \ref{fig:usa_graphs}, \ref{fig:uk_graphs} and \ref{fig:china_graphs} we see the piecewise linear estimation of ARC score using average weighted network distance for cities in the US, UK/Ireland and China respectively. In each case, we see clear and distinct patterns, as discussed in Section IV-B of the main report.

\end{document}